\documentclass[aps,prc,superscriptaddress,showpacs,floatfix, nofootinbib,preprintnumbers,twocolumn]{revtex4}
\usepackage{array}
\usepackage{epsfig}
\usepackage{amsmath}
\usepackage{multirow}
\usepackage{graphicx}
\usepackage{amsmath}
\usepackage{feynmf}
\usepackage{comment}
\usepackage[normalem]{ulem}  
\usepackage[dvips]{color} 

\renewcommand\sout{\bgroup \color{red} \ULdepth=-.5ex \ULset}

\newcommand{\Ex}[2]{\ifmmode{#1\times10^{#2}}\else{$#1\times10^{#2}$}\fi}

\begin{document}
\title{Wilson coefficients of dimension 6 gluon operators with spin indices in the heavy quark current correlation functions}

\author{HyungJoo Kim}\affiliation{Department of Physics and Institute of Physics and Applied Physics, Yonsei
University, Seoul 03722, Korea}
\author{Su Houng Lee}\affiliation{Department of Physics and Institute of Physics and Applied Physics, Yonsei
University, Seoul 03722, Korea}
\date{\today}
\begin{abstract}
 We calculate the Wilson coefficients of dimension 6 gluon operators with spin indices in the two point correlation function of the heavy scalar, pseudoscalar, and axialvector currents. Our result completes the list of all Wilson coefficients of gluon operators up to dimension 6 for the correlation functions between heavy quark currents without derivatives. 
We then use the result to investigate the stability of the QCD sum rule results  for the  $\eta_c$, $J/\Psi$, $\chi_{c0}$, and $\chi_{c1}$ mesons near $T_c$.
While the inclusion of the dimension 6 operators increases the stability of the sum rules for all currents, all of them break down slightly above $T_c$  with that for the $J/\psi$ persisting to relatively highest temperature.  
\end{abstract}


\maketitle

\section{Introduction}
Ever since the seminal work by Matsui and Satz\cite{Matsui:1986dk} that the $J/\psi$ suppression in heavy ion collisions could serve as a signature for the formation of the quark-gluon plasma, a great deal of works have been devoted to the subject\cite{Andronic:2015wma}.  
On the other hand, the question of whether the charmonium states  disappears at $T_c$ or at higher temperature is still controversial up to this date\cite{Kim:2015rdi}.  
While it is generally accepted that  $J/\psi$ is composed dominantly of $c \bar{c}$ quark pair interacting within a non relativistic confining potential at zero temperature, it is not clear what effective potential should be used to analyze the  property at finite temperature\cite{Lee:2013dca,Satz:2015jsa}.

A robust non-perturbative method to calculate the properties of charmonium at zero temperature is based on the QCD sum rule approach\cite{Shifman:1978bx,Reinders:1984sr}.  The method has been found to work well in reproducing the masses of the ground state charmonium states, the electromagnetic decay width of the $J/\psi$, and even the mass difference between $\eta_c$ and $J/\psi$ even before experiment.  In a constituent quark model, the mass difference between the $\eta_c$ and $J/\psi$ is due to the color-spin interaction, which phenomenologically is inversely proportional to the masses of the quark and antiquark.  However, it is well known that a naive generalization of the formula fitted to the light quark system and the light heavy quark system underestimates the mass splittings between heavy-heavy quark systems by a factor of 4~\cite{Lee:2007tn}.  The fact that QCD sum rule fits the mass difference indicates that the method captures well the detailed non-perturbative effects involved in the masses of the charmonium states. 
Therefore, if one is able to systematically investigate the temperature modification of the operator product expansion (OPE) starting from the vacuum, one can hope to learn about the modification as well as the disappearance of the charmonium states above the critical temperature.  

In a previous set of works\cite{Morita:2007pt,Morita:2007hv,Morita:2009qk}, it was shown that one can indeed study the gradual change of the OPE at finite temperature as well as the abrupt change near $T_c$.  This was due to the fact that the temperature dependence of dimension 4 operators, the gluon condensate, and the twist-2 gluon operator that newly appears at finite temperature, could be well extracted from the lattice calculation of the energy density and pressure at finite temperature.  
On the other hand, when the temperature dependencies of the OPE were used in the charmonium sum rules, it was found to break down at $T>1.04 T_c$ for the $J/\psi$ and at $T>1.03T_c$ for the $P$-wave states, respectively~\cite{Song:2008bd}.   

It is not clear if the breakdown of the sum rules is due to the melting of the charmonium states or the breakdown of the OPE. To shed light into the problem,  
we have recently studied the effect of including the dimension 6 operators to the moment sum rules for $J/\psi$ at finite temperature.  It was found that their contribution indeed improved the stability to slightly higher temperature\cite{Kim:2015xna}.  In this work, we calculate the Wilson coefficients of dimension 6 gluon operators with spin indices in the two point correlation function of the heavy scalar, pseudoscalar, and axialvector currents. Our result completes the list of all Wilson coefficients of gluon operators up to dimension 6 for heavy quark currents without derivatives. 
We then use the result to investigate the stability of both the moment and Borel QCD sum rule result for the  $\eta_c$, $J/\psi$, $\chi_{c0}$, and $\chi_{c1}$ mesons near $T_c$.
While the inclusion of the dimension 6 operators increases the stability, the sum rules for all currents break down slightly above $T_c$  with that for the $J/\psi$ persisting to relatively highest temperature.  

The paper is organized as follows.  
In section II, we give a summary of the independent dimension 6 gluon operators.  
Section III shows the calculation for the Wilson coefficients of gluon operator with spin indices up to dimension 6 for each currents. In section IV, we  discuss the temperature dependence and the stability for the moment and Borel sum rules for charmonium states near $T_c$. Section IV is devoted to the summary and a possible interpretation of our result. We list moments and Borel transformed Wilson coefficients in Appendix A and B respectively. Appendix C summarizes the forms for the continuum contribution.  

\section{Independent dimension 6 gluon operators}

Here, we summarize the different representations of the dimension 4 and dimension 6 operators with spin 0 and spin 2~\cite{Kim:2015xna}. 

\subsection{dimension 4 operators} 

There are two independent gluon operators at dimension 4.  They can be represented as scalar and spin 2 operators as follows\cite{Klingl:1998sr}.
\begin{eqnarray}
\langle \frac{\alpha_s}{\pi}G_{\mu \nu}^aG_{\mu \nu}^a \rangle  & = & G_0 , \nonumber \\ 
\langle \frac{\alpha_s}{\pi} G_{\mu \alpha}^a  G_{\nu  \alpha}^a|_{ST}  \rangle & = & (u_\mu u_\nu -\frac{1}{4} g_{\mu \nu} )G_2 ,
\end{eqnarray}
where $ST$ means symmetric and traceless indices and $u_\mu=(1,0,0,0)$ is the medium four vector taken specifically to take the form in this work.
These two independent operators can be recast into the color electric and magnetic condensates as follows.
\begin{eqnarray}
G_0 & = & 2 \langle \frac{\alpha_s}{\pi}(B^2-E^2) \rangle, \nonumber \\
G_2 & = & -\frac{2}{3} \langle \frac{\alpha_s}{\pi}(E^2+B^2) \rangle. 
\end{eqnarray}
Using the energy momentum tensor, one obtains the temperature dependencies of these operators from the lattice measurement of the energy density $\epsilon$ and pressure $p$ in pure gauge theory~\cite{Lee:2008xp}.
\begin{eqnarray}
G_0(T) & = & G_0^{\text{vac}}-\frac{8}{11} (\epsilon -3p), \nonumber \\
G_2(T) & = &  -\frac{\alpha_s (T)}{\pi}(\epsilon + p).
\end{eqnarray}
The temperature dependence in full QCD is expected to be not so different from that obtained in pure gauge theory~\cite{Morita:2007hv}.

\subsection{dimension 6 operators}

There are two independent scalar operators, one of which vanishes in pure gauge theory when the equation of motion is used~\cite{Nikolaev:1982rq}.  
The spin 2 gluon operators were categorized\cite{Kim:2000kj,Kim:2015ywa} and their anomalous dimensions were calculated recently\cite{Kim:2015ywa}.  In purge gauge theory, only one gluon operators survive.  Using covariant derivatives acting on the gluon condensates, the spin 0 and spin 2 gluon operators are 
\begin{eqnarray}
\langle\frac{\alpha_s}{\pi} G^a_{\kappa \lambda} G^a_{\kappa \lambda; \mu \mu} \rangle,&\nonumber\\
\langle\frac{\alpha_s}{\pi} G^a_{\kappa \lambda} G^a_{\kappa \lambda; \mu \nu} |_{ST}\rangle & = ( u_\mu u_\nu -\frac{1}{4} g_{\mu \nu} ) X .
\end{eqnarray}
 Introducing a simplified notation $G^3_{\mu \nu} \equiv f^{abc} G^a_{\mu \alpha}G^b_{\alpha \beta} G^c_{\beta \nu}$, one can express the spin 0 and 2 operators as follows:
\begin{eqnarray}
\langle\frac{\alpha_s}{\pi} G^a_{\kappa \lambda} G^a_{\kappa \lambda; \mu \mu} \rangle & = &    2\langle \frac{g\alpha_s}{\pi} G^3_{\mu \mu} \rangle,
\nonumber \\
X & = &   2\langle\frac{g\alpha_s}{\pi} G_3 \rangle\\,
\nonumber
\end{eqnarray}
in the pure gauge theory where $\langle G^3_{\mu\nu}|_{ST}\rangle = ( u_\mu u_\nu -\frac{1}{4} g_{\mu \nu} )G_3$.  We can again express these operators in terms color electric and color magnetic fields.
\begin{eqnarray}
G^3_{\mu \mu} & = & \langle 3BEE -BBB \rangle, \nonumber \\
G_3 & = &  \frac{1}{3} \langle BBB+BEE \rangle, 
\label{ABC}
\end{eqnarray}
where we have abbreviated the triple scalar product in color space $f^{abc} A^a \cdot (B^b \times C^c )$ as ABC for simplicity.  
While the temperature dependence of the dimension 6 gluon operators can be calculated from the lattice using folded Wilson loops in the future, we will assume their temperature dependence from the following approximation.
\begin{eqnarray}
\langle ABC \rangle = N \bigg( \langle A^2 \rangle \langle B^2 \rangle \langle C^2 \rangle \bigg)^{1/2},
\end{eqnarray}
where the normalization factor $N$ is used to reproduce the vacuum value.

\section{Wilson coefficients}
We start from the following two point correlation function for the scalar(S), pseudoscalar(P), vector(V), and axialvector(A) currents ($J=I,i\gamma_5,\gamma_\mu,(q_{\mu}q_{\nu}/q^2-g_{\mu\nu})\gamma^{\nu}\gamma_5$) and classify it according to spin and dimension of operators which occur in the OPE.

\begin{flalign}
\Pi^{J}(q)&=i\int d^{4}x e^{iqx}\langle T\{j^{J}(x)j^{J}(0)\} \rangle \nonumber\\
&=\Pi^{J}_{scalar}(q)+\Pi^{J}_{4,2}(q)+\Pi^{J}_{6,2}(q)+\Pi^{J}_{6,4}(q).
\end{flalign}
Here, $\Pi^{J}_{scalar}(q)$ denotes the contributions from the scalar operators while $\Pi^{J}_{i,j}$ denotes those from operators with spin indices with the the first and second indices showing the dimension and spin of operator, respectively.

The Wilson coefficients for the scalar gluon operators up to dimension 6 were calculated by Nikolaev and Radyushkin \cite{Nikolaev:1982rq}. The Wilson coefficients for dimension 4 non-zero spin gluon operators are carried out in Refs \cite{Klingl:1998sr}. The new results of this work are the Wilson coefficients of the dimension 6 spin 2, 4 gluon operators for scalar, pseudoscalar, and axialvector currents. The Wilson coefficients of dimension 6 operators with spin indices for the vector current were calculated in Ref.~\cite{Kim:2015xna}, but result for the spin 4 operator is larger than ours by a factor 2.

 In this section, we summarized all Wilson coefficients for non-zero spin operators up to dimension 6. For the notations, we essentially follow those in Ref.~\cite{Kim:2015xna}. $J_N=\int_{0}^{1}\frac{dx}{(1+x(1-x)y)^N}$ with $y=Q^2/m^2$ where $m$ is the heavy quark mass and $Q^2=-q^2$. $h$ denotes the heavy qaurk field.
 
\begin{widetext}
Scalar : $j^S=\bar{h}h$
\begin{flalign}
\Pi^{S}_{4,2}(q)&=\frac{1}{Q^4}\{q^{\mu}q^{\nu}\langle \frac{\alpha_s}{\pi}G^{a}_{\sigma\mu}G^{a}_{\sigma\nu} \rangle[\frac{1}{2}+(1-\frac{1}{3}y)J_{1}-\frac{3}{2}J_{2}]\}\\
\Pi^{S}_{6,2}(q)&=\frac{1}{Q^6}q^{\mu}q^{\nu} \lbrace \nonumber\\
&\left\langle \frac{\alpha_s}{\pi}G^{a}_{\kappa\lambda}G^{a}_{\kappa\lambda ;\mu\nu} \right\rangle [\frac{73}{120}-\frac{1}{60}y+(\frac{5}{4}-\frac{1}{8}y)J_{1}-\frac{37}{8}J_2+\frac{47}{12}J_3-\frac{23}{20}J_4]\nonumber\\
&+\left\langle \frac{\alpha_s}{\pi}G^{a}_{\mu\kappa}G^{a}_{\nu\lambda ;\lambda\kappa} \right\rangle [-\frac{377}{360}+\frac{2}{15}y+(-\frac{19}{12}+\frac{5}{24}y)J_1+\frac{31}{8}J_2-\frac{31}{36}J_3-\frac{23}{60}J_4]\nonumber\\
&+\left\langle \frac{\alpha_s}{\pi}G^{a}_{\mu\kappa}G^{a}_{\kappa\lambda ;\lambda\nu} \right\rangle [\frac{93}{40}-\frac{3}{10}y+(\frac{35}{12}-\frac{3}{8}y)J_1-\frac{67}{8}J_2+\frac{11}{4}J_3+\frac{23}{60}J_4] \rbrace\\
\Pi^{S}_{6,4}(q)&=\frac{1}{Q^8}\{q^{\mu}q^{\nu}q^{\kappa}q^{\lambda}\langle \frac{\alpha_s}{\pi}G^{a}_{\kappa\sigma}G^{a}_{\lambda\sigma ;\mu\nu} \rangle \nonumber\\
&[-\frac{23}{9}+(-\frac{74}{15}+\frac{11}{15}y)J_{1}+\frac{81}{5}J_{2}-\frac{106}{9}J_{3}+\frac{46}{15}J_{4}]\}
\label{scalar}
\end{flalign}

Pseudoscalar : $j^P=\bar{h}i \gamma_{5}h$
\begin{flalign}
\Pi^{P}_{4,2}(q)&=\frac{1}{Q^4}\{q^{\mu}q^{\nu}\langle \frac{\alpha_s}{\pi}G^{a}_{\sigma\mu}G^{a}_{\sigma\nu} \rangle[\frac{1}{2}+(\frac{1}{3}-\frac{1}{3}y)J_{1}-\frac{1}{6}J_{2}-\frac{2}{3}J_3]\}\\
\Pi^{P}_{6,2}(q)&=\frac{1}{Q^6}q^{\mu}q^{\nu} \lbrace \nonumber\\
&\left\langle \frac{\alpha_s}{\pi}G^{a}_{\kappa\lambda}G^{a}_{\kappa\lambda ;\mu\nu} \right\rangle [\frac{41}{120}-\frac{1}{60}y+(\frac{1}{4}-\frac{1}{8}y)J_{1}+\frac{3}{8}J_2-\frac{29}{12}J_3+\frac{37}{20}J_4-\frac{2}{5}J_5]
\nonumber\\
&+\left\langle \frac{\alpha_s}{\pi}G^{a}_{\mu\kappa}G^{a}_{\nu\lambda ;\lambda\kappa} \right\rangle [-\frac{329}{360}+\frac{2}{15}y+(-\frac{11}{12}+\frac{5}{24}y)J_1+\frac{29}{24}J_2+\frac{65}{36}J_3-\frac{21}{20}J_4-\frac{2}{15}J_5]
\nonumber\\
&+\left\langle \frac{\alpha_s}{\pi}G^{a}_{\mu\kappa}G^{a}_{\kappa\lambda ;\lambda\nu} \right\rangle [\frac{263}{120}-\frac{3}{10}y+(\frac{35}{12}-\frac{3}{8}y)J_1-\frac{185}{24}J_2+\frac{25}{12}J_3+\frac{23}{60}J_4+\frac{2}{15}J_5] \rbrace
\\
\Pi^{P}_{6,4}(q)&=\frac{1}{Q^8}\{q^{\mu}q^{\nu}q^{\kappa}q^{\lambda}\langle \frac{\alpha_s}{\pi}G^{a}_{\kappa\sigma}G^{a}_{\lambda\sigma ;\mu\nu} \rangle \nonumber\\
&[-\frac{23}{9}+(-\frac{58}{15}+\frac{11}{15}y)J_{1}+\frac{179}{15}J_{2}-\frac{242}{45}J_{3}-\frac{6}{5}J_{4}+\frac{16}{15}J_{5}]\}
\label{scalar}
\end{flalign}

Vector : $j^V_\mu=\bar{h}\gamma_{\mu}h$
\begin{flalign}
\Pi^{V}_{4,2}|_{\mu\nu}(q)&=\frac{1}{Q^2}[I^{4,2}_{\mu\nu}+\frac{1}{Q^2}(q_{\rho}q_{\mu}I^{4,2}_{\rho\nu}+q_{\rho}q_{\nu}I^{4,2}_{\rho\mu})+g_{\mu\nu}\frac{q_{\rho}q_{\sigma}}{Q^2}J^{4,2}_{\rho\sigma}+\frac{q_{\mu}q_{\nu}q_{\rho}q_{\sigma}}{Q^4}(I^{4,2}_{\rho\sigma}+J^{4,2}_{\rho\sigma})] \\
\Pi^{V}_{6,2}|_{\mu\nu}(q)&=\frac{1}{Q^4}[I^{6,2}_{\mu\nu}+\frac{1}{Q^2}(q_{\rho}q_{\mu}I^{6,2}_{\rho\nu}+q_{\rho}q_{\nu}I^{6,2}_{\rho\mu})+g_{\mu\nu}\frac{q_{\rho}q_{\sigma}}{Q^2}J^{6,2}_{\rho\sigma}+\frac{q_{\mu}q_{\nu}q_{\rho}q_{\sigma}}{Q^4}(I^{6,2}_{\rho\sigma}+J^{6,2}_{\rho\sigma})] \\
\Pi^{V}_{6,4}|_{\mu\nu}(q)&=\frac{q_{\kappa}q_{\lambda}}{Q^6}[I^{6,4}_{\kappa\lambda\mu\nu}+\frac{1}{Q^2}(q_{\rho}q_{\mu}I^{6,4}_{\kappa\lambda\rho\nu}+q_{\rho}q_{\nu}I^{6,4}_{\kappa\lambda\rho\mu})+g_{\mu\nu}\frac{q_{\rho}q_{\sigma}}{Q^2}J^{6,4}_{\kappa\lambda\rho\sigma}+\frac{q_{\mu}q_{\nu}q_{\rho}q_{\sigma}}{Q^4}(I^{6,4}_{\kappa\lambda\rho\sigma}+J^{6,4}_{\kappa\lambda\rho\sigma})]
\label{scalar}
\end{flalign}

\begin{flalign}
I^{4,2}_{\mu\nu}\text{   }=&\left\langle \frac{\alpha_s}{\pi}G^{a}_{\sigma\mu}G^{a}_{\sigma\nu} \right\rangle[\frac{1}{2}+(1-\frac{1}{3}y)J_1-\frac{3}{2}J_2] \\
J^{4,2}_{\mu\nu}\text{   }=&\left\langle \frac{\alpha_s}{\pi}G^{a}_{\sigma\mu}G^{a}_{\sigma\nu} \right\rangle[-\frac{7}{6}+(1+\frac{1}{3}y)J_1-\frac{1}{2}J_2+\frac{2}{3}J_3]\\
I^{6,2}_{\mu\nu}\text{   }=&\left\langle \frac{\alpha_s}{\pi}G^{a}_{\kappa\lambda}G^{a}_{\kappa\lambda ;\mu\nu} \right\rangle[\frac{31}{240}-\frac{1}{60}y+(\frac{13}{24}+\frac{1}{48}y)J_1-\frac{115}{48}J_2+\frac{21}{8}J_3-\frac{9}{10}J_4]\\&
+\left\langle \frac{\alpha_s}{\pi}G^{a}_{\mu\kappa}G^{a}_{\nu\lambda ;\lambda\kappa} \right\rangle[-\frac{739}{720}+\frac{2}{15}y+(-\frac{9}{8}+\frac{3}{16}y)J_1+\frac{133}{48}J_2+\frac{1}{72}J_3-\frac{19}{30}J_4]\\&
+\left\langle \frac{\alpha_s}{\pi}G^{a}_{\mu\kappa}G^{a}_{\kappa\lambda ;\lambda\nu} \right\rangle[\frac{293}{240}-\frac{3}{10}y+(\frac{55}{24}+\frac{1}{16}y)J_1-\frac{131}{16}J_2+\frac{145}{24}J_3-\frac{41}{30}J_4]\\
J^{6,2}_{\mu\nu}\text{   }=&\left\langle \frac{\alpha_s}{\pi}G^{a}_{\kappa\lambda}G^{a}_{\kappa\lambda ;\mu\nu} \right\rangle[\frac{103}{240}+\frac{1}{60}y+(\frac{5}{24}-\frac{7}{48}y)J_1-\frac{59}{48}J_2+\frac{31}{24}J_3-\frac{11}{10}J_4+\frac{2}{5}J_5]\\&
+\left\langle \frac{\alpha_s}{\pi}G^{a}_{\mu\kappa}G^{a}_{\nu\lambda ;\lambda\kappa} \right\rangle[\frac{71}{240}-\frac{2}{15}y+(-\frac{1}{8}+\frac{1}{48}y)J_1+\frac{61}{48}J_2+-\frac{61}{24}J_3+\frac{29}{30}J_4+\frac{2}{15}J_5]\\&
+\left\langle \frac{\alpha_s}{\pi}G^{a}_{\mu\kappa}G^{a}_{\kappa\lambda ;\lambda\nu} \right\rangle[\frac{29}{240}+\frac{3}{10}y+(-\frac{1}{24}-\frac{7}{16}y)J_1+\frac{31}{48}J_2-\frac{23}{24}J_3+\frac{11}{30}J_4-\frac{2}{15}J_5]\\
I^{6,4}_{\mu\nu\rho\sigma}=&\left\langle \frac{\alpha_s}{\pi}G^{a}_{\rho\kappa}G^{a}_{\sigma\kappa ;\mu\nu} \right\rangle[-\frac{133}{45}+(-\frac{10}{3}+\frac{11}{15}y)J_1+\frac{69}{5}J_2-\frac{458}{45}J_3
+\frac{8}{3}J_4]\\
J^{6,4}_{\mu\nu\rho\sigma}=&\left\langle \frac{\alpha_s}{\pi}G^{a}_{\rho\kappa}G^{a}_{\sigma\kappa ;\mu\nu} \right\rangle[\frac{181}{45}-(2+\frac{11}{15}y)J_1-\frac{47}{15}J_2-\frac{22}{45}J_3+\frac{8}{3}J_4-\frac{16}{15}J_5]
\end{flalign}

Axialvector : $j^A_\mu=\bar{h}(q_{\mu}q_{\nu}/q^2-g_{\mu\nu})\gamma_{\nu}\gamma^5 h$
\begin{flalign}
\Pi^{A}_{4,2}|_{\mu\nu}(q)&=\frac{1}{Q^2}[K^{4,2}_{\mu\nu}+\frac{1}{Q^2}(q_{\rho}q_{\mu}K^{4,2}_{\rho\nu}+q_{\rho}q_{\nu}K^{4,2}_{\rho\mu})+g_{\mu\nu}\frac{q_{\rho}q_{\sigma}}{Q^2}L^{4,2}_{\rho\sigma}+\frac{q_{\mu}q_{\nu}q_{\rho}q_{\sigma}}{Q^4}(K^{4,2}_{\rho\sigma}+L^{4,2}_{\rho\sigma})] \\
\Pi^{A}_{6,2}|_{\mu\nu}(q)&=\frac{1}{Q^4}[K^{6,2}_{\mu\nu}+\frac{1}{Q^2}(q_{\rho}q_{\mu}K^{6,2}_{\rho\nu}+q_{\rho}q_{\nu}K^{6,2}_{\rho\mu})+g_{\mu\nu}\frac{q_{\rho}q_{\sigma}}{Q^2}L^{6,2}_{\rho\sigma}+\frac{q_{\mu}q_{\nu}q_{\rho}q_{\sigma}}{Q^4}(K^{6,2}_{\rho\sigma}+L^{6,2}_{\rho\sigma})] \\
\Pi^{A}_{6,4}|_{\mu\nu}(q)&=\frac{q_{\kappa}q_{\lambda}}{Q^6}[K^{6,4}_{\kappa\lambda\mu\nu}+\frac{1}{Q^2}(q_{\rho}q_{\mu}K^{6,4}_{\kappa\lambda\rho\nu}+q_{\rho}q_{\nu}K^{6,4}_{\kappa\lambda\rho\mu})+g_{\mu\nu}\frac{q_{\rho}q_{\sigma}}{Q^2}L^{6,4}_{\kappa\lambda\rho\sigma}+\frac{q_{\mu}q_{\nu}q_{\rho}q_{\sigma}}{Q^4}(K^{6,4}_{\kappa\lambda\rho\sigma}+L^{6,4}_{\kappa\lambda\rho\sigma})]
\label{scalar}
\end{flalign}

\begin{flalign}
K^{4,2}_{\mu\nu}\text{   }=&\left\langle \frac{\alpha_s}{\pi}G^{a}_{\sigma\mu}G^{a}_{\sigma\nu} \right\rangle[\frac{1}{2}+(-1-\frac{1}{3}y)J_1+\frac{1}{2}J_2] \\
L^{4,2}_{\mu\nu}\text{   }=&\left\langle \frac{\alpha_s}{\pi}G^{a}_{\sigma\mu}G^{a}_{\sigma\nu} \right\rangle[-\frac{7}{6}+(\frac{1}{3}+\frac{1}{3}y)J_1+\frac{5}{6}J_2]\\
K^{6,2}_{\mu\nu}\text{   }=&\left\langle \frac{\alpha_s}{\pi}G^{a}_{\kappa\lambda}G^{a}_{\kappa\lambda ;\mu\nu} \right\rangle[-\frac{49}{240}-\frac{1}{60}y+(-\frac{17}{24}+\frac{1}{48}y)J_1+\frac{113}{48}J_2-\frac{43}{24}J_3+\frac{7}{20}J_4]\\&
+\left\langle \frac{\alpha_s}{\pi}G^{a}_{\mu\kappa}G^{a}_{\nu\lambda ;\lambda\kappa} \right\rangle[-\frac{259}{720}+\frac{2}{15}y+(-\frac{3}{8}+\frac{3}{16}y)J_1+\frac{25}{48}J_2-\frac{29}{72}J_3+\frac{37}{60}J_4]\\&
+\left\langle \frac{\alpha_s}{\pi}G^{a}_{\mu\kappa}G^{a}_{\kappa\lambda ;\lambda\nu} \right\rangle[-\frac{9}{80}-\frac{3}{10}y+(-\frac{35}{24}+\frac{1}{16}y)J_1+\frac{81}{16}J_2-\frac{31}{8}J_3+\frac{23}{60}J_4]\\
L^{6,2}_{\mu\nu}\text{   }=&\left\langle \frac{\alpha_s}{\pi}G^{a}_{\kappa\lambda}G^{a}_{\kappa\lambda ;\mu\nu} \right\rangle[\frac{13}{80}+\frac{1}{60}y+(\frac{11}{24}-\frac{7}{48}y)J_1+\frac{1}{48}J_2-\frac{31}{24}J_3+\frac{13}{20}J_4]\\&
+\left\langle \frac{\alpha_s}{\pi}G^{a}_{\mu\kappa}G^{a}_{\nu\lambda ;\lambda\kappa} \right\rangle[\frac{103}{240}-\frac{2}{15}y+(\frac{19}{24}+\frac{1}{48}y)J_1-\frac{103}{48}J_2+\frac{7}{8}J_3+\frac{1}{20}J_4]\\&
+\left\langle \frac{\alpha_s}{\pi}G^{a}_{\mu\kappa}G^{a}_{\kappa\lambda ;\lambda\nu} \right\rangle[-\frac{1}{80}+\frac{3}{10}y+(\frac{17}{24}-\frac{7}{16}y)J_1-\frac{15}{16}J_2+\frac{5}{8}J_3-\frac{23}{60}J_4]\\
K^{6,4}_{\mu\nu\rho\sigma}=&\left\langle \frac{\alpha_s}{\pi}G^{a}_{\rho\kappa}G^{a}_{\sigma\kappa ;\mu\nu} \right\rangle[-\frac{133}{45}+(\frac{2}{3}+\frac{11}{15}y)J_1+\frac{9}{5}J_2+\frac{82}{45}J_3
-\frac{4}{3}J_4]\\
L^{6,4}_{\mu\nu\rho\sigma}=&\left\langle \frac{\alpha_s}{\pi}G^{a}_{\rho\kappa}G^{a}_{\sigma\kappa ;\mu\nu} \right\rangle[\frac{181}{45}+(-\frac{14}{15}-\frac{11}{15}y)J_1-\frac{37}{5}J_2+\frac{266}{45}J_3-\frac{8}{5}J_4]
\end{flalign}
\end{widetext}

\section{Applications : QCD sum rules for Charmonium states near $T_c$}
We use the Wilson coefficient calculation to construct QCD sum rule to investigate the mass of the four charmonium states, $\eta_c$, $J/\Psi$, $\chi_{c0}$, and $\chi_{c1}$, from the two point function of the corresponding P,V,S, and A currents, near $T_c$. 
The moment sum rule for $J/\psi$ was analyzed in our previous paper \cite{Kim:2015xna}.  Here, we generalize the calculation to all charmonium states and also construct and analyze the corresponding Borel sum rule.  
We follow closely our previous paper \cite{Kim:2015xna} to estimate temperature dependence of dimension 6 condensates except for the dimension 6 spin 4 operator $\langle \frac{\alpha_s}{\pi}G^a_{\mu\kappa} G^a_{\nu\kappa;\alpha\beta}|_{ST} \rangle$.  Previously, we estimate this by using gluon distribution function, thermal gluon mass, and Bose distribution function as follows,
\begin{align}
&{G_4}/{G_2} \sim -\left(m_G^2\frac{A_4}{A_2}\right)\frac{{\left\langle p^2\right\rangle}_T}{{\left\langle p^2\right\rangle}_{T_c}},
\end{align}
where $G_4$ is a medium projected operator of $\langle \frac{\alpha_s}{\pi}G^a_{\mu\kappa} G^a_{\nu\kappa;\alpha\beta}|_{ST} \rangle$ defined in Appendix A. But, we neglect ${\left\langle p^2\right\rangle}_T / {\left\langle p^2\right\rangle}_{T_c}$ term in this work  for simplicity and because their contribution are negligible as they are proportional to higher moments of the twist-2 distribution.

\subsection{Moment Sum Rule}
\begin{figure}[h]
  \centering
  \includegraphics[width=0.4\textwidth]{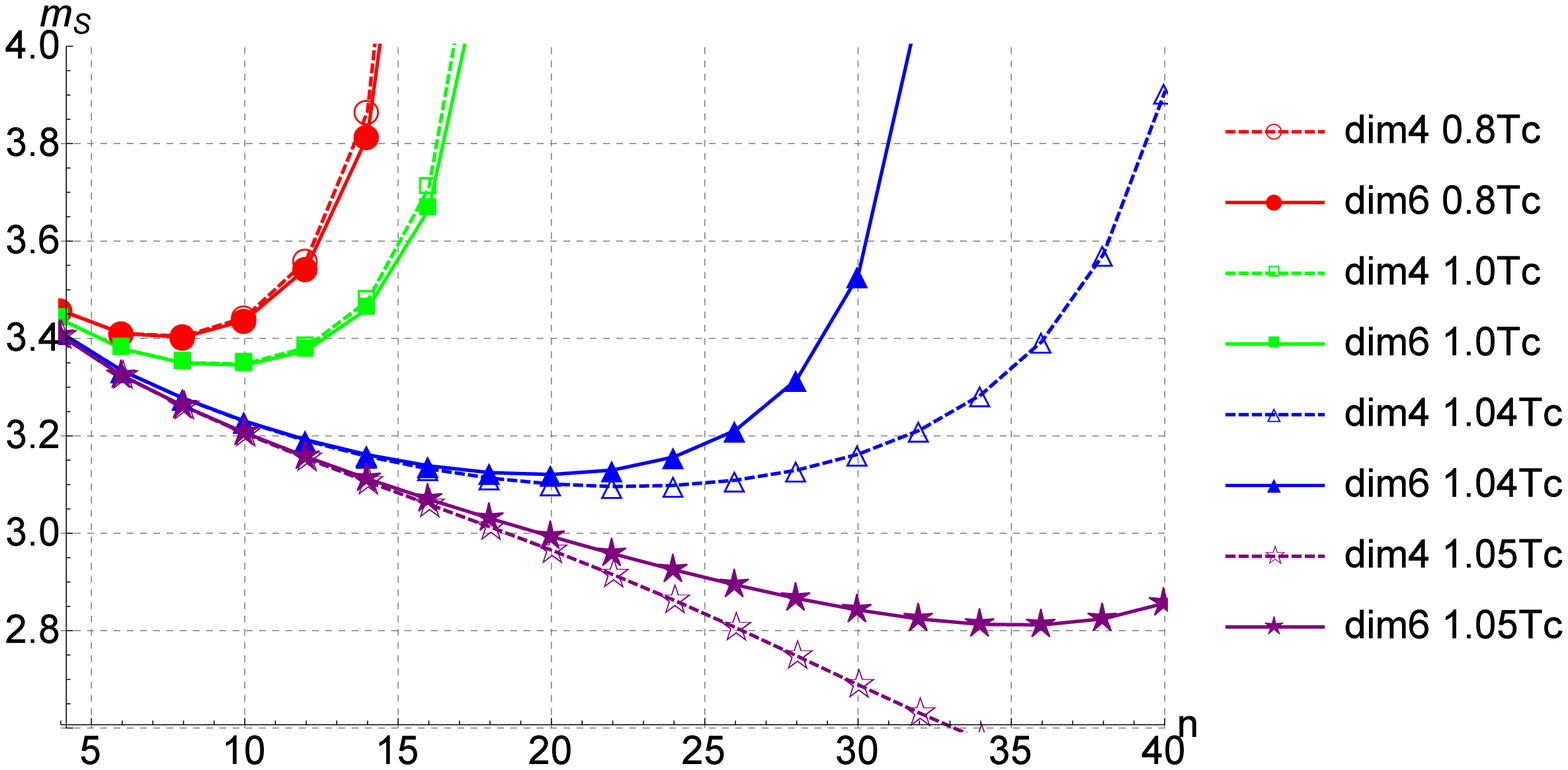}
  \includegraphics[width=0.4\textwidth]{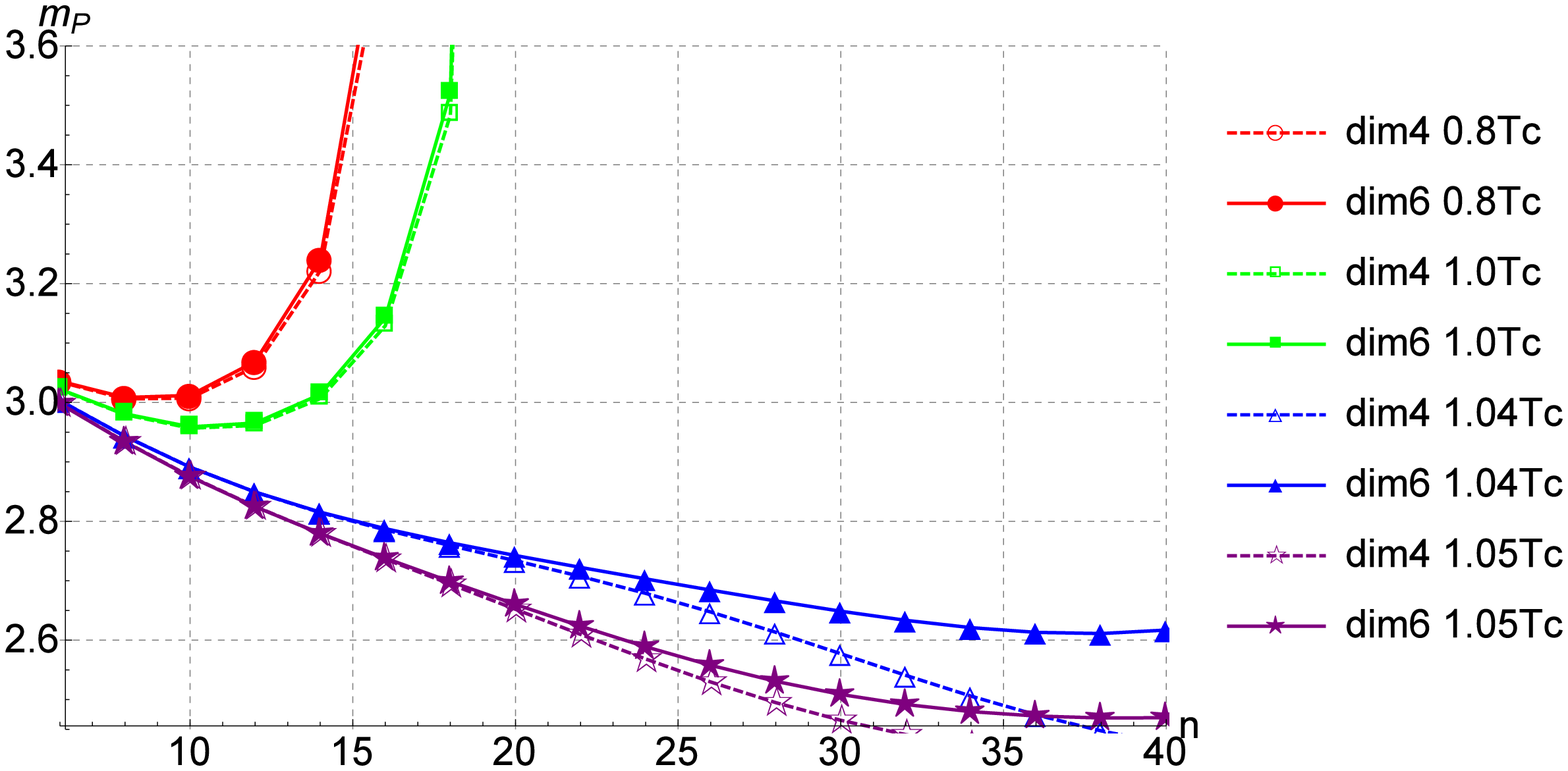}\\
  \includegraphics[width=0.4\textwidth]{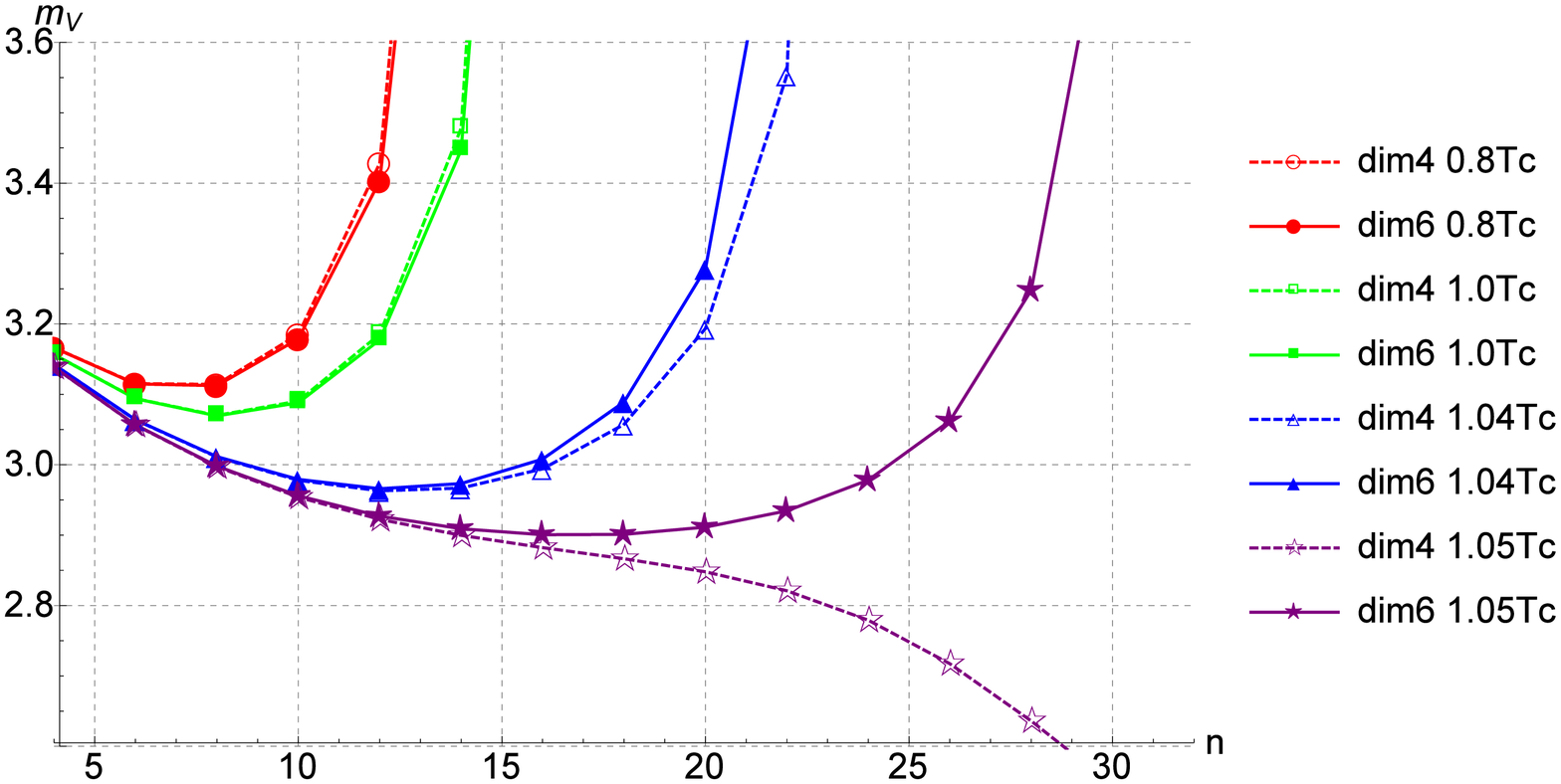}
  \includegraphics[width=0.4\textwidth]{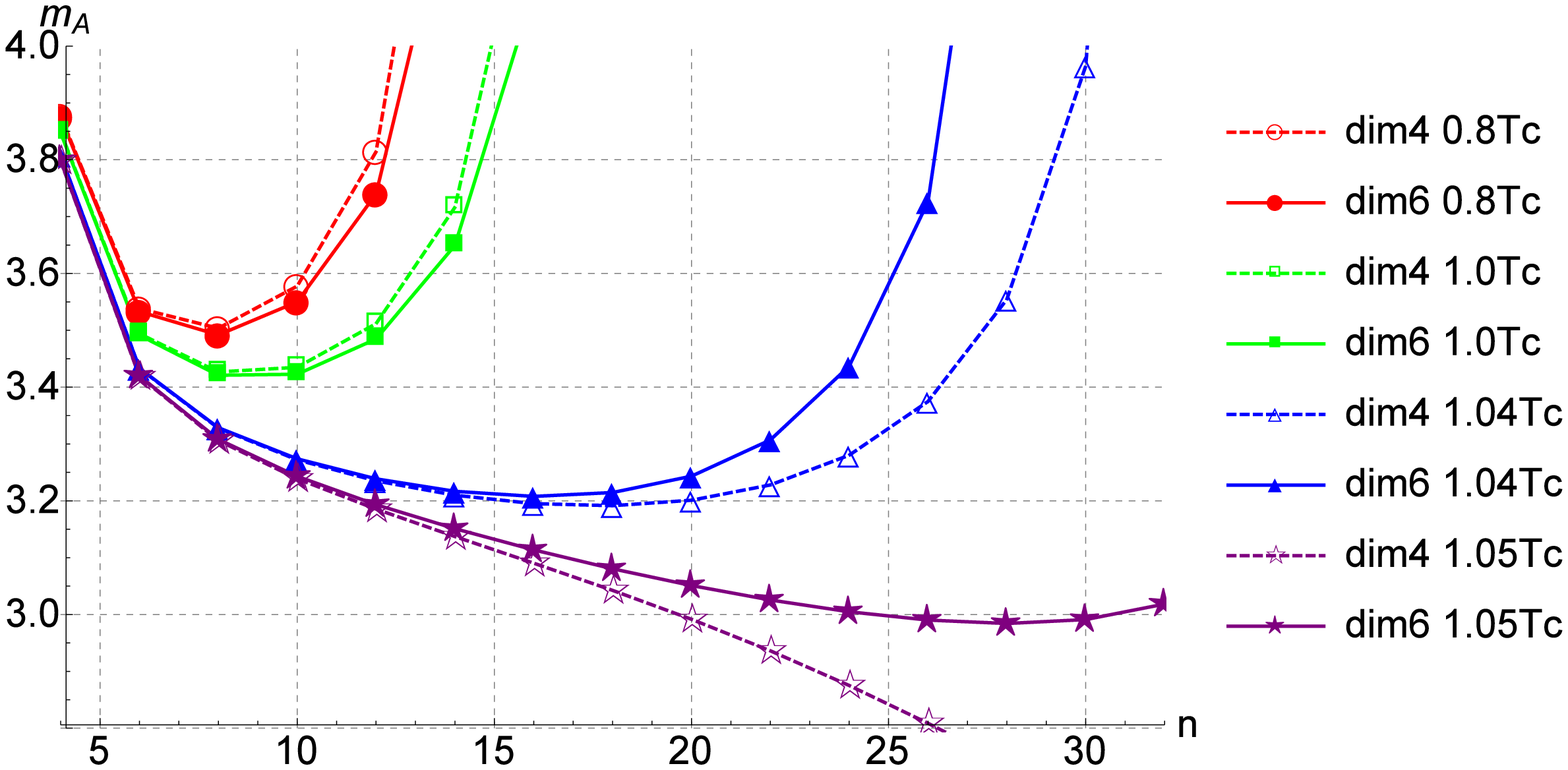}\\
  \caption{Results of moment Sum rules}
  \label{fig1}
\end{figure}

For the first application, we define power moments $M^J_n$ from the correlation function to pick out information of the lowest lying resonance.  By defining the dimensionless correlation function  $\tilde{\Pi}^J$ as  $\tilde{\Pi}^{V,A}={\Pi_{\mu}^{\mu,V,A}}/{(-3q^2)}$ and $\tilde{\Pi}^{S,P}={\Pi^{S,P}}/{q^2}$ for the respective quantum numbers, the moments are given as
\begin{eqnarray}
M^J_n(Q_0^2) =\frac{1}{n!} {\left(-\frac{d}{dQ^2}\right)}^n \tilde{\Pi}^J(Q^2)|_{Q^2=Q_0^2}.
\end{eqnarray}
Assuming that the imaginary part of the correlation function is dominated by very sharp pole from the lowest lying resonance, the charmonium mass is approximately obtained by the extremum value in moment (Fig.~(\ref{fig1})) plot of the following charmonium mass 
\begin{eqnarray}
m_J= \sqrt{\frac{M^J_{n-1}}{M^J_n} -4m_c^2} .
\label{moment}
\end{eqnarray}
For the moment sum rule, we used different parameter sets for S-wave and P-wave reffering to Ref \cite{Reinders:1984sr}. For the continuum, we use the simple form valid in the infinitely heavy quark limit and listed in Appendix C.

\begin{table}[ht]
\centering
\caption{Parameter sets for Moment sum rules}
\begin{tabular}{ |c|c|c|c|c| }
 \hline
         & $\xi(=\frac{Q_0^2}{4m_c^2})$ & $m_c$ [GeV] & $\alpha_s$ &$\sqrt{s}_0$ [GeV]\\
 \hline
 S-waves & 1 & 1.23&0.21&3.8 \\ 
 \hline
 P-waves & 2.5 & 1.21&0.17&3.8 \\ 
 \hline
\end{tabular}
\end{table}

When Eq.~(\ref{moment}) is plotted as a function of $n$, an extremum will appear if the modeling of the imaginary part is consistent with the OPE.  At larger $n$ the plot will deviate from the mass as the truncation of the OPE will fail while  at smaller $n$ the approximation of the continuum will dominate.  The existence of the extremum signals that the mass is reliable and the asympotic expansion is well approximated in both the imaginary part and the OPE part in that region, making the extracted value for the mass reliable.    We determine the reliable  region by imposing the continuum contribution to be less than 30$\%$ of the total perturbative contribution and the dimension 6 contribution to be less than 10$\%$ of the total contribution. In this criterion, we find that $n_{min}$ is about 2$\sim$3 for every current and not important in this work. Only $n_{max}$ at 1.05$T_c$ is critical in determining stability of the sum rules. For S,P,V, and A currents, $n_{max}$ at 1.05$T_c$ is about 25,29,20, and 22 respectively. Thus, we find that only vector particle is stable up to 1.05$T_c$ from the moment sum rule.  Moreover, the extremum exists only for the vector channel. 

According to the sequential charmonium dissociation scenario \cite{Karsch:2005nk}, ground state of psuedoscalar particle might be more stable than other P-wave particles. However, moment sum rule results show that pseudoscalar particle has the worst stability among the four states even though we include dimension 6 contribution. This is  contrary to our expectation.  However, as we will discuss later, it could be a consequence of the approximation of the current sum rule approach as the contribution from disconnected diagrams, which we have neglected, will be large in the PS channel. \\

\subsection{Borel Sum Rule}
To pick up the lowest resonance properties better than from the  moment sum rule, we apply Borel transformation to the previous defined correlation function as follows.  
\begin{equation}
 \mathcal{M}^J(\sigma)  =\lim_{\substack{n/Q^2 \rightarrow \sigma, \\ n,Q^2
  \rightarrow \infty}}
  \frac{(Q^2)^{n+1}\pi}{n!}\left(-\frac{d}{dQ^2}\right)^n
  \tilde{\Pi}^J(Q^2).
\end{equation}
The Borel sum rule is expected to further suppress the continuum contribution than in the moment sum rule even for the heavy quark system\cite{Morita:2009qk}.
For the parameters, we used $m_c$=1.28 GeV and $\alpha_s=0.3$ following Ref \cite{Bertlmann:1981he}. We also include the $\alpha_s$ correction to the continuum contribution as calculated in Ref.~\cite{Reinders:1984sr} and used in Ref.~\cite{Morita:2009qk}, which we list in Appendix C. For the continuum threshold $\bar{s}_0$, we  chose it to make the Borel curve most flat at each temperature. Additionally,  we  introduce the Borel window bounded by $\sigma_{min}$ and $\sigma_{max}$, which are determined using similar criteria to fix the range of $n$ in the case of moment sum rule. The  effective thresholds and Borel windows determined as described above are listed in Table \ref{table:eff-threshold} and \ref{table:borelwindow}, respectively.

\begin{table}[h]
\caption{Effective thresholds}
\label{table:eff-threshold}
\centering
\begin{tabular}{ |c|c|c|c|c| }
 \hline
         & $0.8T_c$ & $1.03T_c$& $1.4T_c$ &$1.05T_c$\\
 \hline
 $\sqrt{\bar{s}_0^S}$& 3.9 & 3.5&3.4&3.1 \\ 
 \hline
 $\sqrt{\bar{s}_0^P}$&3.3 & 3.1&3.0&3.0 \\ 
 \hline
  $\sqrt{\bar{s}_0^V}$&3.5 & 3.2&3.1&3.1 \\ 
 \hline
  $\sqrt{\bar{s}_0^A}$& 3.8 & 3.5&3.4&3.0 \\ 
 \hline
\end{tabular}
\caption{Borel windows with dimension 6 contribution}
\label{table:borelwindow}
\centering
\begin{tabular}{ |c|c c|c c|c c|c c| }
 \hline
 $T/T_c$&$\sigma^{P}_{min}$&$\sigma^{P}_{max}$&$\sigma^{V}_{min}$&$\sigma^{V}_{max}$&$\sigma^{S}_{min}$&$\sigma^{S}_{max}$&$\sigma^{A}_{min}$&$\sigma^{A}_{max}$\\
 \hline
 0.8&0.38&1.38&0.26&1.9&0.26&0.81&0.26&0.69\\
 \hline
1.03&0.62&2.1&0.47&3.2&0.56&1.2&0.40&2.0\\
\hline
1.04&0.89&2.2&0.62&1.6&0.56&1.2&0.47&1.1\\
\hline
1.05&0.89&2.4&0.62&1.6&1.1&1.5&0.9&1.0\\
\hline
\end{tabular}
\label{borelwindow}
\end{table}

\begin{figure}[h]
  \centering
  \includegraphics[width=0.4\textwidth]{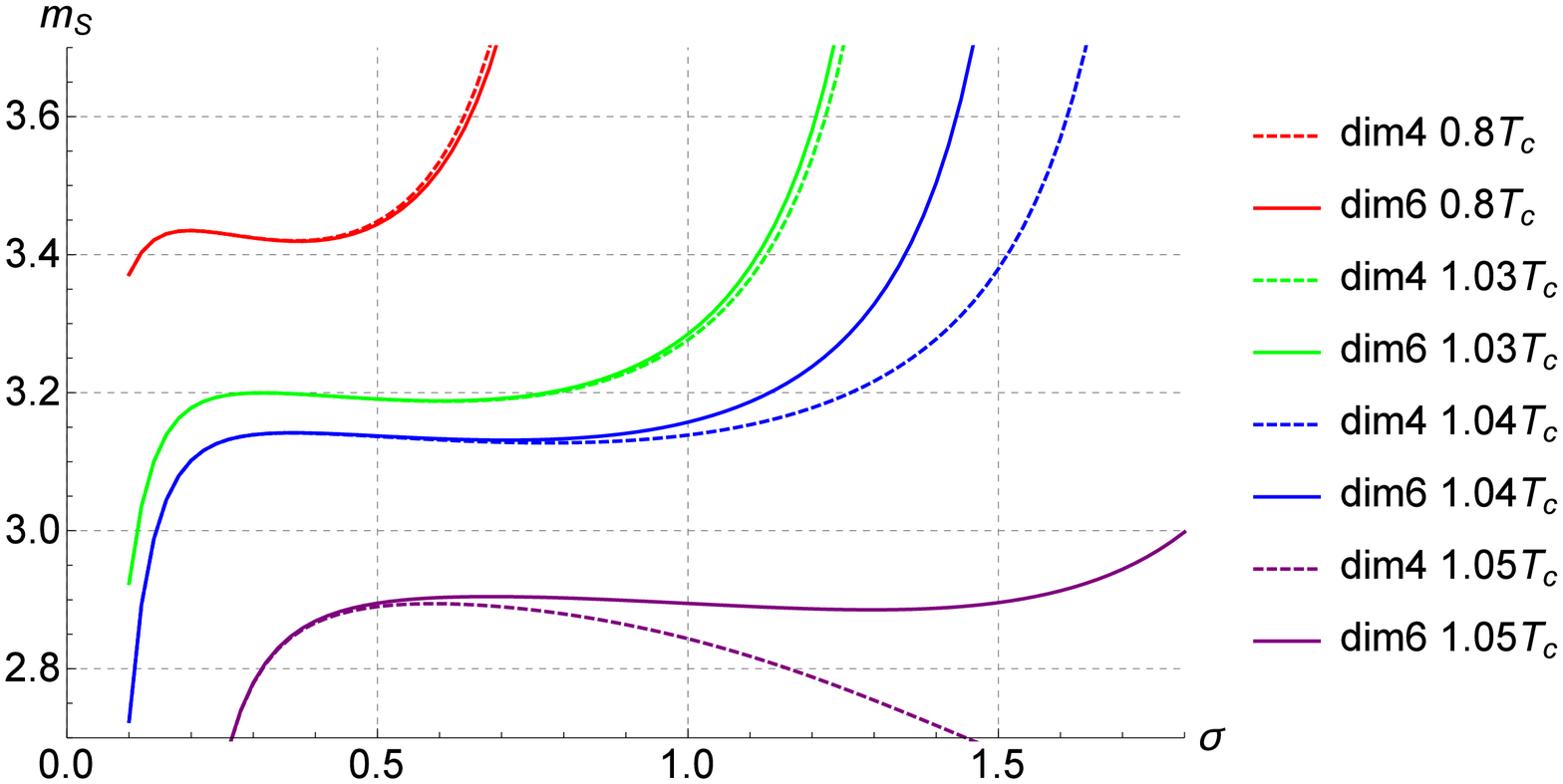}
  \includegraphics[width=0.4\textwidth]{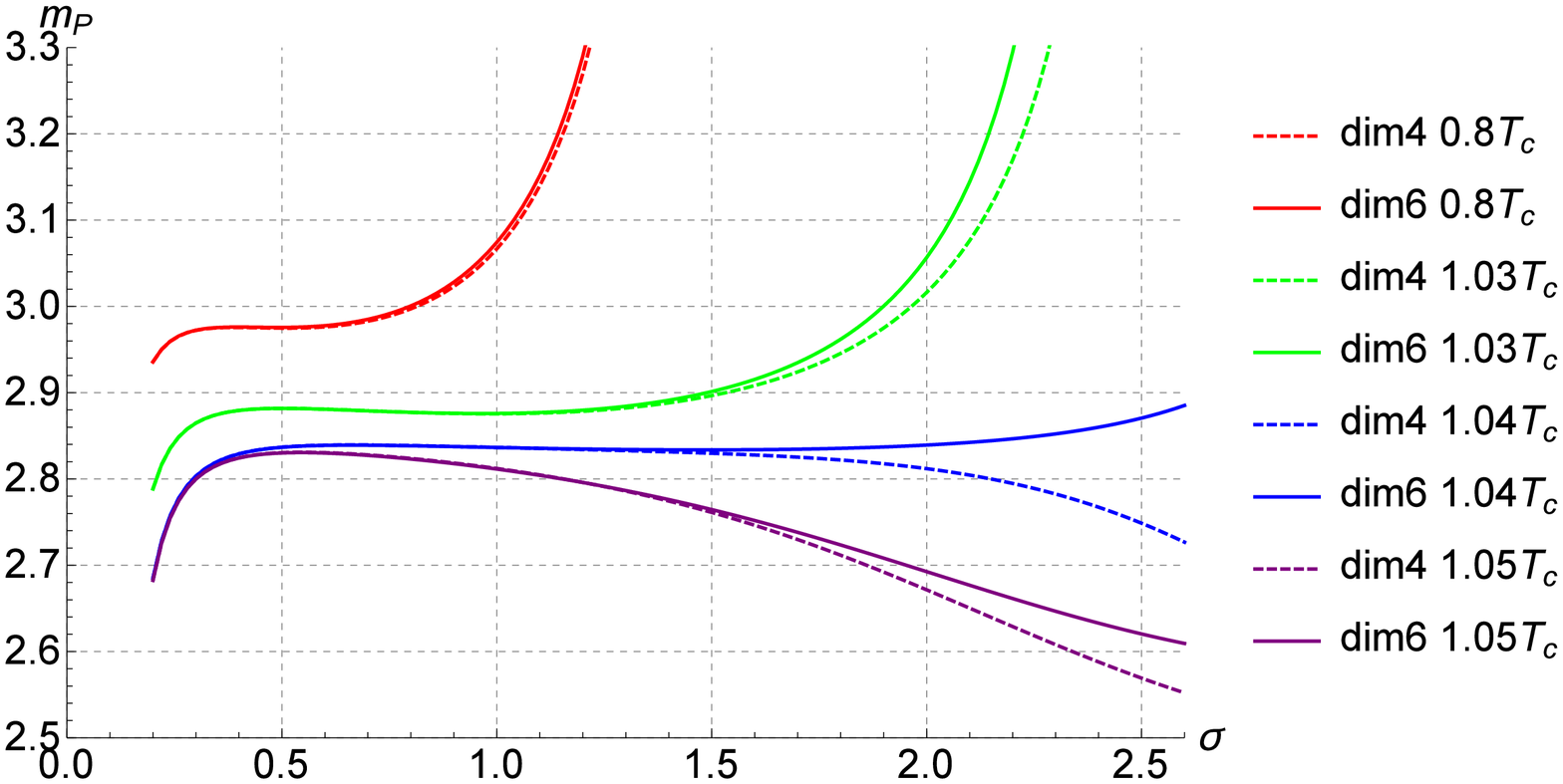}\\
  \includegraphics[width=0.4\textwidth]{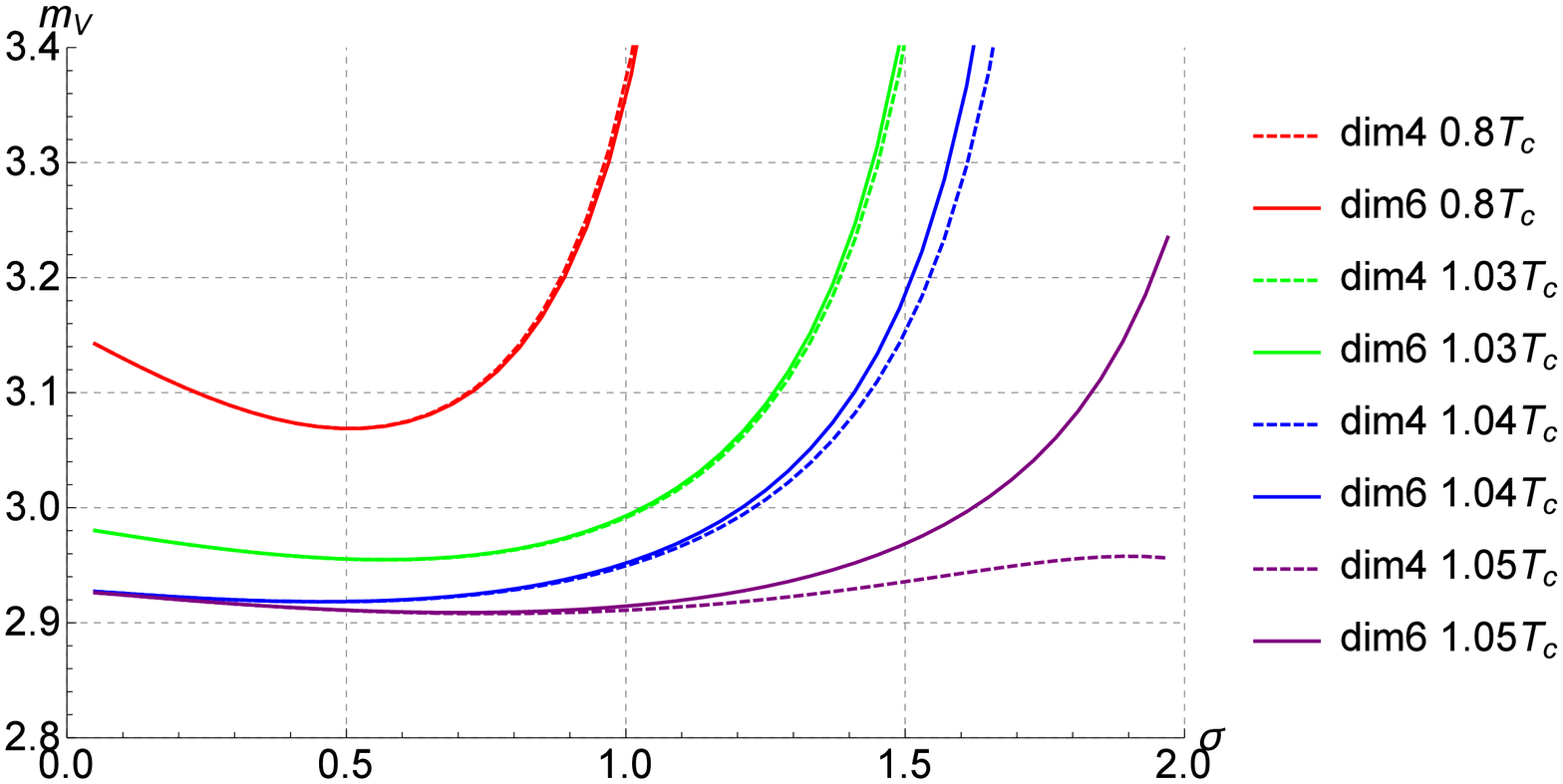}
  \includegraphics[width=0.4\textwidth]{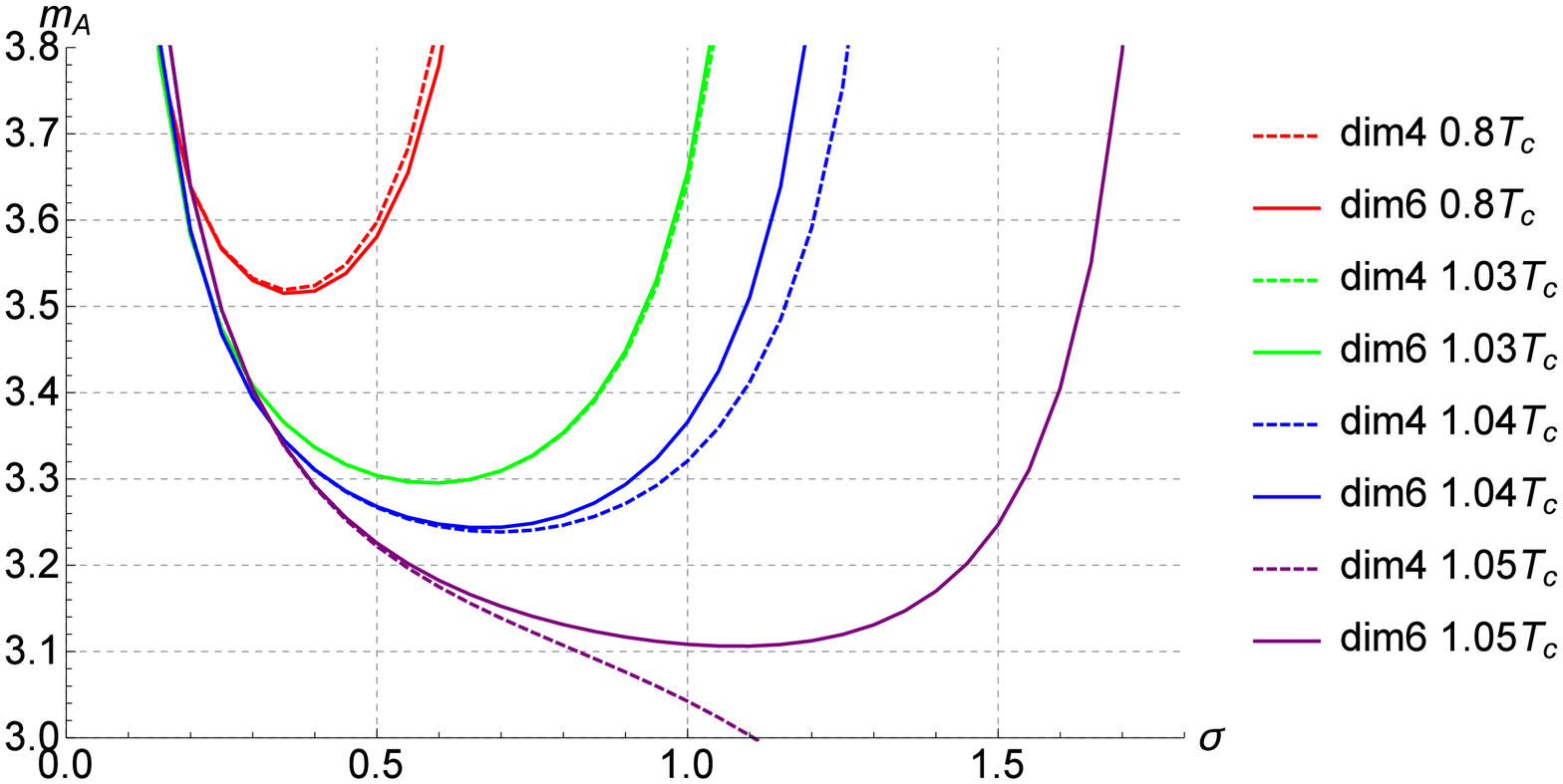}\\
  \caption{Results of Borel Sum rules with temperature dependent threshold with  $\alpha_s$ correction.}
  \label{borel}
\end{figure}

As can be seen from Fig.~(\ref{borel}), the Borel curve for the vector mass shows stability up to 1.05$T_c$ with a wide Borel window consistent with the results from the moment sum rule. The scalar and axialvector particles has a minimum in the Borel curve up to 1.05$T_c$ but the Borel windows at this temperature are too narrow (Table \ref{borelwindow}).  Thus the sum rules seem stable only up to 1.04 $T_c$. Pseudoscalar particle has a wide Borel window at 1.05$T_c$ but has no stability.  This suggests that a sharp pole ansatz is inconsistent with the OPE or there is something missing in the OPE.  

It is useful to look at the result in connection with the natural decay widths of these states as given in Table~\ref{width}.  The states we are considering can not decay into open charm hadrons.  Hence, their hadronic decays are dominated by the disconnected diagrams that decay into light hadrons.  The fact that the vacuum width of $\eta_c$ is large is an experimental proof that the disconnected diagrams are large in this channel.    
Similar situation is also manifest in the light quark sector as seen by the 
small mixing angle and ideal mixing of the flavor octet and singlet representation in the pseudoscalar and the vector channel, which are  consequences of the large and small contributions of the disconnected diagrams in the respective channels.    This means that in the pseudoscalar channel, the OPE should have a nontrivial contribution from the disconnected diagrams, which we do not include.  It also suggests that the large width in the vacuum will induce an even larger increase in the width at finite temperature, which should be properly taken into account at finite temperatures for this channel. 
A thermal width in the sum rule will slightly change the temperature dependence of the $J/\psi$ mass near $T_c$~\cite{Morita:2009qk}.   
However, a more detailed study allowing for a more realistic imaginary part that takes into account various form for the thermal width and scattering effects as well as an estimate of the disconnected diagrams should be a topic to pursed in the future.  
We summarized mass, decay width, and maximum temperature which stability of sum rule is maintaind for both moment and Borel sum rule for each current state in Table \ref{relativestability}.

\begin{table}[h]
\caption{Relative stability and vacuum width}
\label{relativestability}
\centering
\begin{scriptsize}
\begin{tabular}{ |c|c|c|c|c| }
 \hline
         State& Mass &Width& Moment & Borel\\
 \hline
P($\eta_c$)& 2.98GeV& 31.8MeV&$\leq$ 1.03$T_c$&$\leq$ 1.04$T_c$ \\ 
 \hline
 V($J/\Psi$)&3.09GeV&92.9keV&$\leq$ 1.05$T_c$&$\leq$ 1.05$T_c$ \\ 
 \hline
S($\chi_{c0}$)& 3.41GeV&10.5MeV& $\leq$ 1.04$T_c$&$\lesssim$ 1.05$T_c$(narrow window)\\ 
 \hline
A($\chi_{c1}$)& 3.51GeV&0.84MeV& $\leq$ 1.04$T_c$&$\lesssim$ 1.05$T_c$(narrow window)\\ 
 \hline
\end{tabular}
\end{scriptsize}
\label{width}
\end{table}

\section{Summary}

 We have presented the Wilson coefficients of dimension 6 gluon operators with spin indices in the two point correlation function of the heavy scalar, pseudoscalar, and axialvector currents. Our result completes the list of all Wilson coefficients of gluon operators up to dimension 6 for the correlation functions between heavy quark currents without derivatives. 
We have also presented results on  the stability of the QCD sum rule results  for the  $\eta_c$, $J/\Psi$, $\chi_{c0}$, and $\chi_{c1}$ mesons near $T_c$.
While the inclusion of the dimension 6 operators increases the stability, the sum rules for all currents break down slightly above $T_c$  with that for the $J/\psi$ persisting to relatively highest temperature.   As our approach is  based on a systematic extension of the vacuum QCD sum rules for the charmonium states, the results should be reliable up to the temperature where  the stability persist.  Above the temperature, the OPE convergence as well as the naive peak structure of the spectral densities becomes questionable.  
One possible interpretation of our result could be that all the charmonium states dissolves at slightly above the critical temperature or at least that  a naive peak structures with width are not consistent with the OPE above these temperatures.  To confirm these results, it would be useful to work out the dimension 8 contribution as it is known that dimension 6 contribution has a smaller effect compared to dimension 4 and 8 operators due to to the extra suppressing numerical factor in the Wilson coefficients~\cite{Nikolaev:1982rq}. Also, it would be useful to link the operators contributing dominantly in the OPE representation of the space-time Wilson loops to those contributing to the two point correlation functions.  This together with a more realistic modeling of the imaginary part above $T_c$ will be a useful future research topic to understand the real dissolving temperatures of the charmonium states.  
\section*{Appendix A: Moments}
In this Appendix, we list all moments for completeness.
\begin{small}
\begin{align}
M^{J}_n(\xi)=&A^{J}_n(\xi)[1+a^{J}_n(\xi)\alpha_s+b^{J}_n(\xi)\phi^4_b+c^{J}_n(\xi)\phi^4_c \nonumber\\
&+s^{J}_n(\xi)\phi^6_s+t^{J}_n(\xi)\phi^6_t+x^{J}_n(\xi)\phi^6_x \nonumber\\
&+y^{J}_n(\xi)\phi^6_y+z^{J}_n(\xi)\phi^6_z+g^{J}_{4n}(\xi)\phi^6_{g_4}]
\end{align}
\end{small}
\begin{small}
\begin{align}
\phi^4_b=&\frac{4\pi^2}{9}\frac{G_0 }{(4m^2)^2}\\
\phi^4_c=&\frac{4\pi^2}{3}\frac{G_2}{(4m^2)^2}&\\
\phi^6_s=&\frac{4\pi^2}{3\cdot 1080}\frac{\langle\frac{\alpha_s}{\pi}G^a_{\mu\nu}G^a_{\mu\nu;\kappa\kappa}\rangle}{(4m^2)^3}\\
\phi^6_t=&\frac{2}{3}\frac{4\pi^2}{3\cdot 1080}\frac{\langle\frac{\alpha_s}{\pi}G^a_{\alpha\kappa}G^a_{\alpha\lambda;\lambda\kappa}\rangle}{(4m^2)^3}\\
\phi^6_x=&\frac{9}{2}\frac{4\pi^2}{3\cdot 1080}\frac{X}{(4m^2)^3}\\
\phi^6_y=&\frac{3}{2}\frac{4\pi^2}{3\cdot 1080}\frac{Y}{(4m^2)^3}\\
\phi^6_z=&\frac{3}{2}\frac{4\pi^2}{3\cdot 1080}\frac{Z}{(4m^2)^3}\\
\phi^6_{g_4}=&5\frac{4\pi^2}{3\cdot 1080}\frac{G_4}{(4m^2)^3}
\end{align}
\end{small}
We project out the non-zero spin operators using the medium four velocity $u_{\mu}=(1,0,0,0)$. 
\begin{small}
\begin{align}
\left\langle \frac{\alpha_s}{\pi}G^a_{\sigma\mu} G^a_{\sigma\nu} |_{ST}\right\rangle & =(u_\mu u_\nu -\frac{1}{4}g_{\mu\nu})G_2\nonumber\\
\left\langle \frac{\alpha_s}{\pi}G^a_{\kappa\lambda} G^a_{\kappa\lambda;\mu\nu}|_{ST} \right\rangle & =(u_\mu u_\nu -\frac{1}{4}g_{\mu\nu})X\nonumber\\
 \left\langle \frac{\alpha_s}{\pi}G^a_{\mu\kappa} G^a_{\nu\lambda;\lambda\kappa}|_{ST} \right\rangle & =(u_\mu u_\nu -\frac{1}{4}g_{\mu\nu})Y\nonumber\\
 \left\langle \frac{\alpha_s}{\pi}G^a_{\mu\kappa} G^a_{\kappa\lambda;\lambda\nu}|_{ST} \right\rangle & =(u_\mu u_\nu -\frac{1}{4}g_{\mu\nu})Z\nonumber\\
\left\langle \frac{\alpha_s}{\pi}G^a_{\mu\kappa} G^a_{\nu\kappa;\alpha\beta}|_{ST} \right\rangle & =(u_\mu u_\nu u_\alpha u_\beta+\frac{1}{48}(g_{\mu\nu}g_{\alpha\beta}+g_{\mu\alpha}g_{\nu\beta} \nonumber\\
&+g_{\mu\beta} g_{\nu\alpha}) -\frac{1}{8}(u_\mu u_\nu g_{\alpha\beta}+u_\mu u_\alpha g_{\nu\beta}\nonumber\\
&+u_\mu u_\beta g_{\alpha\nu}+u_\nu u_\alpha g_{\mu\beta} + u_\nu u_\beta g_{\mu\alpha} \nonumber\\
&+ u_\alpha u_\beta g_{\mu\nu}))G_4.
\end{align}
\end{small}
\begin{scriptsize}
\begin{align}
f_n(c_1,...,c_k)=&c_1(n+3)\,_2F_1(n+1,\frac{1}{2},n+\frac{9}{2},\rho)\nonumber\\
&-2c_2(2n+7)\,_2F_1(n+1,\frac{1}{2},n+\frac{7}{2},\rho)\nonumber\\
&+\Sigma_{j=3}^{k}c_k\frac{(n+1+k)!}{(n+2)!(k-2)!}\,_2F_1(n+1,\frac{5}{2}-k,n+\frac{9}{2},\rho)
\end{align}
\end{scriptsize}

Scalar:
\begin{scriptsize}
\begin{align}
A^S_n(\xi)=&\frac{3}{8\pi^2}\frac{3\cdot 2^n(n-1)!}{(2n+3)!!(4m^2)^n(1+\xi)^n}\,_2F_1(n,\frac{3}{2},n+\frac{5}{2},\rho)\\
\sigma^S_n(\xi)=&\frac{-8}{1+\xi}\frac{n(n+1)(n+2)}{(2n+5)(2n+7)}\\
b^S_n(\xi)=&-\frac{3n(n+1)(n+2)(n+3)}{(2n+5)(1+\xi)}\nonumber\\
&\frac{\,_2F_1(n+1,-\frac{1}{2},n+\frac{7}{2},\rho)-\frac{2}{(n+3)}\,_2F_1(n+1,\frac{1}{2},n+\frac{7}{2},\rho)}{\,_2F_1(n,\frac{3}{2},n+\frac{5}{2},\rho)}\\
c^S_n(\xi)=&b^S_n(\xi)-\frac{4n(n+1)}{3(1+\xi)}\frac{\,_2F_1(n+1,\frac{1}{2},n+\frac{5}{2},\rho)}{\,_2F_1(n,\frac{3}{2},n+\frac{5}{2},\rho)}\\
s^S_n(\xi)=&\sigma^S_n(\xi)\frac{f_n(90,0,-420,440,-138)}{\,_2F_1(n,\frac{3}{2},n+\frac{5}{2},\rho)}\\
t^S_n(\xi)=&s^S_n(\xi)+\sigma^S_n(\xi)\frac{f_n(-480,-60,2220,-1680)}{\,_2F_1(n,\frac{3}{2},n+\frac{5}{2},\rho)}\\
x^S_n(\xi)=&s^S_n(\xi)+\sigma^S_n(\xi)\frac{f_n(60,-15,-135,30)}{\,_2F_1(n,\frac{3}{2},n+\frac{5}{2},\rho)}\\
y^S_n(\xi)=&s^S_n(\xi)+\sigma^S_n(\xi)\frac{f_n(-660,75,1815,-750)}{\,_2F_1(n,\frac{3}{2},n+\frac{5}{2},\rho)}\\
z^S_n(\xi)=&s^S_n(\xi)+\sigma^S_n(\xi)\frac{f_n(960,-135,-2595,550,276)}{\,_2F_1(n,\frac{3}{2},n+\frac{5}{2},\rho)}\\
g^S_{4n}(\xi)=&s^S_n(\xi)+\sigma^S_n(\xi)\frac{f_n(132,-33,-309,90)}{\,_2F_1(n,\frac{3}{2},n+\frac{5}{2},\rho)}
\end{align}
\end{scriptsize}

Pseudoscalar:
\begin{scriptsize}
\begin{align}
A^P_n(\xi)=&\frac{3}{8\pi^2}\frac{3\cdot 2^n(n-1)!}{(2n+1)!!(4m^2)^n(1+\xi)^n}\,_2F_1(n,\frac{1}{2},n+\frac{3}{2},\rho)\\
\sigma^P_n(\xi)=&\frac{-8\cdot 3}{1+\xi}\frac{n(n+1)(n+2)}{(2n+3)(2n+5)(2n+7)}\\
b^P_n(\xi)=&-\frac{n(n+1)(n+2)(n+3)}{(2n+3)(1+\xi)}\nonumber\\
&\frac{\,_2F_1(n+1,-\frac{3}{2},n+\frac{5}{2},\rho)-\frac{6}{(n+3)}\,_2F_1(n+1,-\frac{1}{2},n+\frac{5}{2},\rho)}{\,_2F_1(n,\frac{1}{2},n+\frac{3}{2},\rho)}\\
c^P_n(\xi)=&b^P_n(\xi)-\frac{4n(n+1)}{(1+\xi)}\frac{\,_2F_1(n+1,-\frac{1}{2},n+\frac{3}{2},\rho)}{\,_2F_1(n,\frac{1}{2},n+\frac{3}{2},\rho)}
\\
s^P_n(\xi)=&\sigma^P_n(\xi)\frac{f_n(-90,0,480,-620,306,-48)}{\,_2F_1(n,\frac{1}{2},n+\frac{3}{2},\rho)}\\
t^P_n(\xi)=&s^P_n(\xi)+\sigma^P_n(\xi)\frac{f_n(840,-60,-3420,2920,-720)}{\,_2F_1(n,\frac{1}{2},n+\frac{3}{2},\rho)}\\
x^P_n(\xi)=&s^P_n(\xi)+\sigma^P_n(\xi)\frac{f_n(120,-15,-435,330,-84)}{\,_2F_1(n,\frac{1}{2},n+\frac{3}{2},\rho)}\\
y^P_n(\xi)=&s^P_n(\xi)+\sigma^P_n(\xi)\frac{f_n(-240,75,-45,1270,-684)}{\,_2F_1(n,\frac{1}{2},n+\frac{3}{2},\rho)}\\
z^P_n(\xi)=&s^P_n(\xi)+\sigma^P_n(\xi)\frac{f_n(1140,-135,-3255,1370,-168,96)}{\,_2F_1(n,\frac{1}{2},n+\frac{3}{2},\rho)}\\
g^P_{4n}(\xi)=&s^P_n(\xi)+\sigma^P_n(\xi)\frac{f_n(264,-33,-1017,862,-252)}{\,_2F_1(n,\frac{1}{2},n+\frac{3}{2},\rho)}
\end{align}
\end{scriptsize}

Vector:
\begin{scriptsize}
\begin{align}
A^V_n(\xi)=&\frac{3}{4\pi^2}\frac{2^n(n+1)(n-1)!}{(2n+3)!!(4m^2)^n(1+\xi)^n}\,_2F_1(n,\frac{1}{2},n+\frac{5}{2},\rho)\\
\sigma^V_n(\xi)=&\frac{-8}{1+\xi}\frac{n(n+2)}{(2n+5)(2n+7)}\\
b^V_n(\xi)=&-\frac{n(n+1)(n+2)(n+3)}{(2n+5)(1+\xi)^2}\frac{\,_2F_1(n+2,-\frac{1}{2},n+\frac{7}{2};\rho)}{\,_2F_1(n,\frac{1}{2},n+\frac{5}{2};\rho)}\\
c^V_n(\xi)=&b^V_n(\xi)-\frac{4n(n+1)}{3(2n+5)(1+\xi)^2}\frac{\,_2F_1(n+2,\frac{3}{2},n+\frac{7}{2};\rho)}{\,_2F_1(n,\frac{1}{2},n+\frac{5}{2};\rho)}\\
s^V_n(\xi)=&\sigma^V_n(\xi)\frac{f_n(0,0,120,-310,258,-72)}{\,_2F_1(n,\frac{1}{2},n+\frac{5}{2};\rho)}\\
t^V_n(\xi)=&s^V_n(\xi)+\sigma^V_n(\xi)\frac{f_n(-120,-60,-300,1320,-720)}{\,_2F_1(n,\frac{1}{2},n+\frac{5}{2};\rho)}\\
x^V_n(\xi)=&s^V_n(\xi)+\sigma^V_n(\xi)\frac{f_n(-70,25,245,-80,-6)}{\,_2F_1(n,\frac{1}{2},n+\frac{5}{2};\rho)}\\
y^V_n(\xi)=&s^V_n(\xi)+\sigma^V_n(\xi)\frac{f_n(270,-45,-1305,1680,-666)}{\,_2F_1(n,\frac{1}{2},n+\frac{5}{2};\rho)}\\
z^V_n(\xi)=&s^V_n(\xi)+\sigma^V_n(\xi)\frac{f_n(-390,225,1005,-260,-210,144)}{\,_2F_1(n,\frac{1}{2},n+\frac{5}{2};\rho)}\\
g^V_{4n}(\xi)=&s^V_n(\xi)+\sigma^V_n(\xi)\frac{f_n(-210,-33,-444,48,-18)}{\,_2F_1(n,\frac{1}{2},n+\frac{5}{2};\rho)}
\end{align}
\end{scriptsize}

Axialvector:
\begin{scriptsize}
\begin{align}
A^A_n(\xi)=&\frac{2}{3}A^S_n(\xi)\\
\sigma^A_n(\xi)=&\frac{-8}{1+\xi}\frac{n(n+1)(n+2)}{(2n+5)(2n+7)}\\
b^A_n(\xi)=&-\frac{3n(n+1)(n+2)(n+3)}{(2n+5)(1+\xi)}\frac{\,_2F_1(n+1,-\frac{1}{2},n+\frac{7}{2};\rho)}{\,_2F_1(n,\frac{3}{2},n+\frac{5}{2};\rho)}\\
c^A_n(\xi)=&b^A_n(\xi)+\frac{4n(n+1)}{3(1+\xi)}\frac{\,_2F_1(n+1,\frac{1}{2},n+\frac{5}{2};\rho)}{\,_2F_1(n,\frac{3}{2},n+\frac{5}{2};\rho)}\\
s^A_n(\xi)=&\sigma^A_n(\xi)\frac{f_n(0,0,-150,250,-138)}{\,_2F_1(n,\frac{3}{2},n+\frac{5}{2};\rho)}\\
t^A_n(\xi)=&s^A_n(\xi)+\sigma^A_n(\xi)\frac{f_n(-120,-60,1140,-1080)}{\,_2F_1(n,\frac{3}{2},n+\frac{5}{2};\rho)}\\
x^A_n(\xi)=&s^A_n(\xi)+\sigma^A_n(\xi)\frac{f_n(-40,25,5,50)}{\,_2F_1(n,\frac{3}{2},n+\frac{5}{2};\rho)}\\
y^A_n(\xi)=&s^A_n(\xi)+\sigma^A_n(\xi)\frac{f_n(-360,-45,1215,-650)}{\,_2F_1(n,\frac{3}{2},n+\frac{5}{2};\rho)}\\
z^A_n(\xi)=&s^A_n(\xi)+\sigma^A_n(\xi)\frac{f_n(-120,225,-255,70,276)}{\,_2F_1(n,\frac{3}{2},n+\frac{5}{2};\rho)}\\
g^A_{4n}(\xi)=&s^A_n(\xi)+\sigma^A_n(\xi)\frac{f_n(-48,-33,-309,150)}{\,_2F_1(n,\frac{3}{2},n+\frac{5}{2};\rho)}
\end{align}
\end{scriptsize}

\section*{Appendix B: Exponential Moments}
In this Appendix, we list Borel transformed Wilson coefficients for dimension 6 condensates.  The expressions for $\phi^6_i$. $A^{J}_{\omega}$, $a^{J}_{\omega}$, $b^{J}_{\omega}$, and $c^{J}_{\omega}$ are listed in \cite{Bertlmann:1981he,Morita:2009qk}. $G(a,b,\omega)$ is the Whittaker function.
\begin{small}
\begin{align}
\mathcal{M}^{J}(\omega)=&e^{-\omega}\pi A^{J}_{\omega}[1+a^{J}_{\omega}\alpha_s+b^{J}_{\omega}\phi^4_b+c^{J}_{\omega}\phi^4_c+s^{J}_{\omega}\phi^6_s \nonumber\\
&+t^{J}_{\omega}\phi^6_t+x^{J}_{\omega}\phi^6_x+y^{J}_{\omega}\phi^6_y+z^{J}_{\omega}\phi^6_z+g^{J}_{4\omega}\phi^6_{g_4}]
\end{align}
\end{small}
Scalar:
\begin{scriptsize}
\begin{align}
s^S_{\omega }=&\frac{2 \omega}{G\left(\frac{3}{2},\frac{5}{2},\omega \right)}\bigg{(}-90 G\left(\frac{1}{2},\frac{7}{2},\omega \right)+420 G\left(-\frac{1}{2},\frac{7}{2},\omega \right)\nonumber\\
&-220 G\left(-\frac{3}{2},\frac{7}{2},\omega \right)+23 G\left(-\frac{5}{2},\frac{7}{2},\omega \right)\bigg{)}\\
t^S_{\omega }=&\frac{120 \omega}{G\left(\frac{3}{2},\frac{5}{2},\omega \right)}
\bigg{(}-4 G\left(\frac{1}{2},\frac{5}{2},\omega \right)+8 G\left(\frac{1}{2},\frac{7}{2},\omega \right)\nonumber\\
&-37 G\left(-\frac{1}{2},\frac{7}{2},\omega \right)+14 G\left(-\frac{3}{2},\frac{7}{2},\omega \right)\bigg{)}\\
x^S_{\omega }=&-\frac{30 \omega}{G\left(\frac{3}{2},\frac{5}{2},\omega \right)}\bigg{(}4 G\left(\frac{1}{2},\frac{5}{2},\omega \right)+4 G\left(\frac{1}{2},\frac{7}{2},\omega \right)\nonumber\\
&-9 G\left(-\frac{1}{2},\frac{7}{2},\omega \right)+G\left(-\frac{3}{2},\frac{7}{2},\omega \right)\bigg{)}\\
y^S_{\omega }=&\frac{30 \omega}{G\left(\frac{3}{2},\frac{5}{2},\omega \right)}\bigg{(}20 G\left(\frac{1}{2},\frac{5}{2},\omega \right)+44 G\left(\frac{1}{2},\frac{7}{2},\omega \right)\nonumber\\
&-121 G\left(-\frac{1}{2},\frac{7}{2},\omega \right)+25 G\left(-\frac{3}{2},\frac{7}{2},\omega \right)\bigg{)}\\
z^S_{\omega }=&-\frac{2 \omega}{G\left(\frac{3}{2},\frac{5}{2},\omega \right)}\bigg{(}540 G\left(\frac{1}{2},\frac{5}{2},\omega \right)+960 G\left(\frac{1}{2},\frac{7}{2},\omega \right)\nonumber\\
&-2595 G\left(-\frac{1}{2},\frac{7}{2},\omega \right)+275 G\left(-\frac{3}{2},\frac{7}{2},\omega \right)+46 G\left(-\frac{5}{2},\frac{7}{2},\omega \right)\bigg{)}\\
g^S_{4\omega }=&-\frac{6 \omega}{G\left(\frac{3}{2},\frac{5}{2},\omega \right)}\bigg{(}44 G\left(\frac{1}{2},\frac{5}{2},\omega \right)+44 G\left(\frac{1}{2},\frac{7}{2},\omega \right)\nonumber\\
&-103 G\left(-\frac{1}{2},\frac{7}{2},\omega \right)+15 G\left(-\frac{3}{2},\frac{7}{2},\omega \right)\bigg{)}
\end{align}
\end{scriptsize}

Pseudoscalar:
\begin{scriptsize}
\begin{align}
s^P_{\omega }=&\frac{3 \omega}{G\left(\frac{1}{2},\frac{3}{2},\omega \right)}\bigg{(}90 G\left(\frac{1}{2},\frac{7}{2},\omega \right)-480 G\left(-\frac{1}{2},\frac{7}{2},\omega \right)\nonumber\\
&+310 G\left(-\frac{3}{2},\frac{7}{2},\omega \right)-51 G\left(-\frac{5}{2},\frac{7}{2},\omega \right)+2 G\left(-\frac{7}{2},\frac{7}{2},\omega \right)\bigg{)}\\
t^P_{\omega }=&\frac{60 \omega  }{G\left(\frac{1}{2},\frac{3}{2},\omega \right)}\bigg{(}-12 G\left(\frac{1}{2},\frac{5}{2},\omega \right)-42 G\left(\frac{1}{2},\frac{7}{2},\omega \right)\nonumber\\
&+171 G\left(-\frac{1}{2},\frac{7}{2},\omega \right)-73 G\left(-\frac{3}{2},\frac{7}{2},\omega \right)+6 G\left(-\frac{5}{2},\frac{7}{2},\omega \right)\bigg{)}\\
x^P_{\omega }=&\frac{3 \omega  }{G\left(\frac{1}{2},\frac{3}{2},\omega \right)}\bigg{(}-60 G\left(\frac{1}{2},\frac{5}{2},\omega \right)-120 G\left(\frac{1}{2},\frac{7}{2},\omega \right)\nonumber\\
&+435 G\left(-\frac{1}{2},\frac{7}{2},\omega \right)-165 G\left(-\frac{3}{2},\frac{7}{2},\omega \right)+14 G\left(-\frac{5}{2},\frac{7}{2},\omega \right)\bigg{)}\\
y^P_{\omega }=&\frac{3 \omega  }{G\left(\frac{1}{2},\frac{3}{2},\omega \right)}\bigg{(}300 G\left(\frac{1}{2},\frac{5}{2},\omega \right)+240 G\left(\frac{1}{2},\frac{7}{2},\omega \right)\nonumber\\
&+45 G\left(-\frac{1}{2},\frac{7}{2},\omega \right)-635 G\left(-\frac{3}{2},\frac{7}{2},\omega \right)+114 G\left(-\frac{5}{2},\frac{7}{2},\omega \right)\bigg{)}\\
z^P_{\omega }=&-\frac{3 \omega  }{G\left(\frac{1}{2},\frac{3}{2},\omega \right)}\bigg{(}540 G\left(\frac{1}{2},\frac{5}{2},\omega \right)+1140 G\left(\frac{1}{2},\frac{7}{2},\omega \right)\nonumber\\
&-3255 G\left(-\frac{1}{2},\frac{7}{2},\omega \right)+685 G\left(-\frac{3}{2},\frac{7}{2},\omega \right)-28 G\left(-\frac{5}{2},\frac{7}{2},\omega \right)\nonumber\\&+4 G\left(-\frac{7}{2},\frac{7}{2},\omega \right)\bigg{)}\\
g^P_{4\omega }=&\frac{3 \omega  }{G\left(\frac{1}{2},\frac{3}{2},\omega \right)}\bigg{(}-132 G\left(\frac{1}{2},\frac{5}{2},\omega \right)-264 G\left(\frac{1}{2},\frac{7}{2},\omega \right)\nonumber\\
&+1017 G\left(-\frac{1}{2},\frac{7}{2},\omega \right)-431 G\left(-\frac{3}{2},\frac{7}{2},\omega \right)+42 G\left(-\frac{5}{2},\frac{7}{2},\omega \right)\bigg{)}
\end{align}
\end{scriptsize}

Vector:
\begin{scriptsize}
\begin{align}
s^V_{\omega }=&\frac{2 \omega  }{G\left(\frac{1}{2},\frac{5}{2},\omega \right)}\bigg{(}-120 G\left(-\frac{1}{2},\frac{7}{2},\omega \right)+155 G\left(-\frac{3}{2},\frac{7}{2},\omega \right)\nonumber\\
&-43 G\left(-\frac{5}{2},\frac{7}{2},\omega \right)+3 G\left(-\frac{7}{2},\frac{7}{2},\omega \right)\bigg{)}\\
t^V_{\omega }=&\frac{120 \omega  }{G\left(\frac{1}{2},\frac{5}{2},\omega \right)}\bigg{(}-4 G\left(\frac{1}{2},\frac{5}{2},\omega \right)+2 G\left(\frac{1}{2},\frac{7}{2},\omega \right)\nonumber\\
&+5 G\left(-\frac{1}{2},\frac{7}{2},\omega \right)-11 G\left(-\frac{3}{2},\frac{7}{2},\omega \right)+2 G\left(-\frac{5}{2},\frac{7}{2},\omega \right)\bigg{)}\\
x^V_{\omega }=&\frac{2 \omega  }{G\left(\frac{1}{2},\frac{5}{2},\omega \right)}\bigg{(}100 G\left(\frac{1}{2},\frac{5}{2},\omega \right)+70 G\left(\frac{1}{2},\frac{7}{2},\omega \right)\nonumber\\
&-245 G\left(-\frac{1}{2},\frac{7}{2},\omega \right)+40 G\left(-\frac{3}{2},\frac{7}{2},\omega \right)+G\left(-\frac{5}{2},\frac{7}{2},\omega \right)\bigg{)}\\
y^V_{\omega }=&-\frac{6 \omega  }{G\left(\frac{1}{2},\frac{5}{2},\omega \right)}\bigg{(}60 G\left(\frac{1}{2},\frac{5}{2},\omega \right)+90 G\left(\frac{1}{2},\frac{7}{2},\omega \right)\nonumber\\
&-435 G\left(-\frac{1}{2},\frac{7}{2},\omega \right)+280 G\left(-\frac{3}{2},\frac{7}{2},\omega \right)-37 G\left(-\frac{5}{2},\frac{7}{2},\omega \right)\bigg{)}\\
z^V_{\omega }=&\frac{2 \omega  }{G\left(\frac{1}{2},\frac{5}{2},\omega \right)}\bigg{(}900 G\left(\frac{1}{2},\frac{5}{2},\omega \right)+390 G\left(\frac{1}{2},\frac{7}{2},\omega \right)\nonumber\\
&-1005 G\left(-\frac{1}{2},\frac{7}{2},\omega \right)+130 G\left(-\frac{3}{2},\frac{7}{2},\omega \right)+35 G\left(-\frac{5}{2},\frac{7}{2},\omega \right)\nonumber\\
&-6 G\left(-\frac{7}{2},\frac{7}{2},\omega \right)\bigg{)}\\
g^V_{4\omega }=&\frac{6 \omega  }{G\left(\frac{1}{2},\frac{5}{2},\omega \right)}\bigg{(}-44 G\left(\frac{1}{2},\frac{5}{2},\omega \right)+70 G\left(\frac{1}{2},\frac{7}{2},\omega \right)\nonumber\\
&+148 G\left(-\frac{1}{2},\frac{7}{2},\omega \right)-8 G\left(-\frac{3}{2},\frac{7}{2},\omega \right)+G\left(-\frac{5}{2},\frac{7}{2},\omega \right)\bigg{)}
\end{align}
\end{scriptsize}

Axialvector:
\begin{scriptsize}
\begin{align}
s^A_{\omega }=&\frac{2 \omega  }{G\left(\frac{3}{2},\frac{5}{2},\omega \right)}\bigg{(}150 G\left(-\frac{1}{2},\frac{7}{2},\omega \right)-145 G\left(-\frac{3}{2},\frac{7}{2},\omega \right)\nonumber\\
&+23 G\left(-\frac{5}{2},\frac{7}{2},\omega \right)\bigg{)}\\
t^A_{\omega }=&\frac{120 \omega  }{G\left(\frac{3}{2},\frac{5}{2},\omega \right)}\bigg{(}-4 G\left(\frac{1}{2},\frac{5}{2},\omega \right)+2 G\left(\frac{1}{2},\frac{7}{2},\omega \right)\nonumber\\
&-19 G\left(-\frac{1}{2},\frac{7}{2},\omega \right)+9 G\left(-\frac{3}{2},\frac{7}{2},\omega \right)\bigg{)}\\
x^A_{\omega }=&-\frac{10 \omega  }{G\left(\frac{3}{2},\frac{5}{2},\omega \right)}\bigg{(}-20 G\left(\frac{1}{2},\frac{5}{2},\omega \right)-8 G\left(\frac{1}{2},\frac{7}{2},\omega \right)\nonumber\\
&+G\left(-\frac{1}{2},\frac{7}{2},\omega \right)+5 G\left(-\frac{3}{2},\frac{7}{2},\omega \right)\bigg{)}\\
y^A_{\omega }=&\frac{30 \omega  }{G\left(\frac{3}{2},\frac{5}{2},\omega \right)}\bigg{(}-12 G\left(\frac{1}{2},\frac{5}{2},\omega \right)+24 G\left(\frac{1}{2},\frac{7}{2},\omega \right)\nonumber\\
&-81 G\left(-\frac{1}{2},\frac{7}{2},\omega \right)+23 G\left(-\frac{3}{2},\frac{7}{2},\omega \right)\bigg{)}\\
z^A_{\omega }=&-\frac{2 \omega  }{G\left(\frac{3}{2},\frac{5}{2},\omega \right)}\bigg{(}-900 G\left(\frac{1}{2},\frac{5}{2},\omega \right)-120 G\left(\frac{1}{2},\frac{7}{2},\omega \right)\nonumber\\
&-255 G\left(-\frac{1}{2},\frac{7}{2},\omega \right)+35 G\left(-\frac{3}{2},\frac{7}{2},\omega \right)+46 G\left(-\frac{5}{2},\frac{7}{2},\omega \right)\bigg{)}\\
g^A_{4\omega }=&\frac{6 \omega }{G\left(\frac{3}{2},\frac{5}{2},\omega \right)} \bigg{(}-44 G\left(\frac{1}{2},\frac{5}{2},\omega \right)+16 G\left(\frac{1}{2},\frac{7}{2},\omega \right)\nonumber\\
&+103 G\left(-\frac{1}{2},\frac{7}{2},\omega \right)-25 G\left(-\frac{3}{2},\frac{7}{2},\omega \right)\bigg{)}
\end{align}
\end{scriptsize}

\section*{Appendix C: Continuum part of the Spectral function}
We summarize the forms for the continuum part of the spectral function which we use in the Moment and Borel sum rules.
For the Moment sum rule, we use following simple form.
\begin{small}
\begin{align}
\text{Im}\tilde{\Pi}^{J}_{cont}(s)&=(1+\frac{\alpha_s}{\pi})\text{Im}\tilde{\Pi}^{J}_{0}(s)\theta(s-s_0)\\
\text{Im}\tilde{\Pi}^{S}_{0}(s)&=\frac{3u^3}{8\pi}\\
\text{Im}\tilde{\Pi}^{P}_{0}(s)&=\frac{3u}{8\pi}\\
\text{Im}\tilde{\Pi}^{V}_{0}(s)&=\frac{u(3-u^2)}{8\pi}\\
\text{Im}\tilde{\Pi}^{A}_{0}(s)&=\frac{u^3}{4\pi}
\end{align}
\end{small}
For the Bore sum rule, we consider the full $\alpha_s$ correction calculated in Ref.~\cite{Reinders:1984sr} and well summarzied in Ref.~\cite{Morita:2009qk}. 
\begin{small}
\begin{align}
\text{Im}\tilde{\Pi}^{J}_{cont}(s)&=\text{Im}\tilde{\Pi}^{J}(s)\theta(s-s_0)\\
\text{Im}\tilde{\Pi}^{S}(s)&=\text{Im}\tilde{\Pi}^{S}_{0}(s)[1+\frac{4\alpha_s}{3\pi u^3}\{\pi u^3[\frac{\pi}{2u}-\frac{1+u}{2}(\frac{\pi}{2}-\frac{3}{\pi})]\nonumber\\
&+(\frac{15}{16}-\frac{3u^2}{8}-\frac{33u^4}{16})\ln\frac{1+u}{1-u}-\frac{15}{8}u+\frac{33}{8}u^3\}]\nonumber\\
&-\frac{9u(1-u^2)}{8\pi}\Delta\\
\text{Im}\tilde{\Pi}^{P}(s)&=\text{Im}\tilde{\Pi}^{P}_{0}(s)[1+\frac{4\alpha_s}{3 \pi u} \{\pi u[\frac{\pi}{2u}-\frac{3+u}{4}(\frac{\pi}{2}-\frac{3}{4\pi})]
\nonumber\\
&+u-\frac{3(u^6-7u^4+19u^2+3)}{16(3-u^2)} \ln \frac{1+u}{1-u}\nonumber\\
&+\frac{3u(11-4u^2+u^4)}{8(3-u^2)}\}]-\frac{3(1-u^2)}{8\pi u}\Delta\\
\text{Im}\tilde{\Pi}^{V}(s)&=\text{Im}\tilde{\Pi}^{V}_{0}(s)[1+\frac{4\alpha_s}{3}[\frac{\pi}{2u}-\frac{u+3}{4}(\frac{\pi}{2}-\frac{3}{4\pi})]\nonumber\\
&-\frac{3(1-u^2)^2}{8\pi u}\Delta\\
\text{Im}\tilde{\Pi}^{A}(s)&=\text{Im}\tilde{\Pi}^{A}_{0}(s)[1+\frac{4\alpha_s}{3\pi u^3}\{\pi u^3[\frac{\pi}{2u}-\frac{1+u}{2}(\frac{\pi}{2}-\frac{3}{\pi})]\nonumber\\
&+u^3+\frac{3(15-7u^2-7u^4-u^6)}{32}\ln\frac{1+u}{1-u}\nonumber\\
&+\frac{3(u^5-2u^3-15)}{16}    \}]-\frac{3u(1-u^2)}{4\pi}\Delta
\end{align}
\end{small}
where $u=\sqrt{1-\frac{4m^2}{s}}$ and $\Delta=\frac{2\alpha_s}{\pi}\ln 2$.
\section*{Acknowledgements}

This work was supported by the Korea National Research
Foundation under the grant number 2016R1D1A1B03930089.

\end{document}